\documentclass[pre,aps,reprint,amsmath,amsfonts,showpacs,showkeys]{revtex4-1}

\usepackage{algpseudocode}
\usepackage{amsthm}
\usepackage{bm}
\usepackage{color}
\usepackage{fixmath}
\usepackage{graphicx}

\draft 

\theoremstyle{plain}

\DeclareMathOperator{\diag}{diag}
\DeclareMathOperator{\Ord}{O}
\DeclareMathOperator{\tr}{tr}

\newcommand{\Statexi}{\Statex\hspace{\algorithmicindent}}

\begin{document}


\title{The symmetries of image formation by scattering. I. Theoretical framework} 



\author{Dimitrios Giannakis}
\email[All authors contributed equally. Correspondence should be addressed to: ]{ourmazd@uwm.edu}
\affiliation{Courant Institute of Mathematical Sciences, New York University, 251 Mercer St, New York, NY 10012}

\author{Peter Schwander}
\email[All authors contributed equally. Correspondence should be addressed to: ]{ourmazd@uwm.edu}
\author{Abbas Ourmazd}
\email[All authors contributed equally. Correspondence should be addressed to: ]{ourmazd@uwm.edu}
\affiliation{Dept.~of Physics, University of Wisconsin Milwaukee, 1900 E. Kenwood Blvd, Milwaukee, WI 53211}


\date{\today}

\begin{abstract}
We perceive the world through images formed by scattering. The ability to interpret scattering data mathematically has opened to our scrutiny the constituents of matter, the building blocks of life, and the remotest corners of the universe. Here, we deduce for the first time the fundamental symmetries underlying image formation. Intriguingly, these are similar to those of the anisotropic ``Taub universe'' of general relativity, with eigenfunctions closely related to spinning tops in quantum mechanics. This opens the possibility to apply the powerful arsenal of tools developed in two major branches of physics to new problems. We augment these tools with graph-theoretic means to recover the three-dimensional structure of objects from random snapshots of unknown orientation at four orders of magnitude higher complexity than previously demonstrated.  Our theoretical framework offers a potential link to recent observations on face perception in higher primates. In a later paper, we demonstrate the recovery of structure and dynamics from ultralow-signal random sightings of systems with no orientational or timing information.
\end{abstract}

\pacs{02.40.Ky, 61.05.-a, 87.64.Bx, 87.64.Ee, 89.20.Ff, 36.20.-r, 87.15.-v}
\keywords{Structure determination, manifold embedding, differential geometry, X-ray scattering, electron microscopy, tomography, machine learning}

\maketitle 

\section{Introduction}
\label{secIntroduction}

We perceive by constructing three-dimensional (3D) models from random sightings of objects in different orientations. Our ability to recognize that we are seeing a profile even when the face is unknown \cite{FreiwaldTsao10} suggests that perception may rely on object-independent properties of the image formation process. The discovery of these properties would underpin our mathematical formalisms, enhance our computational reach, and perhaps elucidate the process of perception. 

Here, we show that image formation by scattering possesses specific symmetries, which stem from the nature of operations in three-dimensional (3D) space. This follows from a theoretical framework, which considers the information contained in a collection of random sightings of an object. An example is a collection of two-dimensional (2D) snapshots of a moving head, or a rotating molecule. Our primary results can be summarized as follows. (1) The information gleaned from random sightings of an object by scattering onto a 2D detector can be represented as a Riemannian manifold with properties resembling well-known systems in general relativity and quantum mechanics.  As shown below and in a subsequent paper, this allows one to efficiently construct a 3D model of the object. (2) The information about an object can be decomposed into an object-independent term reminiscent of a Platonic Form or ``ideal object'' \cite{Ross51}, plus an object-specific fingerprint. (3) The object-independent term can be used to determine the object orientation giving rise to each snapshot independently of the object, while the object-specific fingerprint can be used for recognition purposes. This seems to have been experimentally observed in face recognition in higher primates \cite{FreiwaldTsao10}.

More generally, it is now well-established that numerical data clouds give rise to manifolds, whose properties can be accessed by powerful graph-theoretic methods  \cite{TenenbaumEtAl00,RoweisSaul00,BelkinNiyogi03,DonohoEtAl03,CoifmanEtAl05,CoifmanLafon06}. However, it has proved difficult to assign physical meaning to the results  \cite{CoifmanEtAl10,FergusonEtAl10}. Using the arsenal of tools developed in differential geometry, general relativity, and quantum mechanics, we elucidate the physical meaning of the outcome of graph-theoretic analysis of scattering data, without the need for restrictive a priori assumptions \cite{SingerEtAl10}. Finally, perception can be formulated as learning some functions on the observation manifold. Our approach is then tantamount to machine learning with a dictionary acquired from the empirically accessible eigenfunctions of well-known operators.

After a brief conceptual outline in Sec.~\ref{secConceptualOutline}, we summarize relevant previous work in Sec.~\ref{PreviousWork}.  The symmetry-based scheme for analyzing scattering data is developed in Sec.~\ref{secTheory}, and the utility of these concepts demonstrated in Sec.~\ref{secResults} in the context of simulated images from X-ray scattering. We present our conclusions in Sec.~\ref{secConclusions}. Material of a more technical nature, including mathematical derivations and algorithms, is provided in Appendices~\ref{appInducedMetric} and~\ref{appAlgorithms}.  A movie of a reconstructed object in 3D is presented as supplementary online material \cite{EPAPS}. Further applications are described in a subsequent paper \cite{GiannakisEtAl11b}, hereafter referred to as Paper~II.   

\section{Conceptual outline}
\label{secConceptualOutline}

We are concerned with constructing a model from sightings of a system viewed in some projection, i.e., by accessing a subset of the variables describing the state of the system. A 3D model of an object, for example, can be constructed from an ensemble of 2D snapshots. Each snapshot can be represented by a vector with the pixel intensities as components. A collection of snapshots then forms a cloud of points in some high-dimensional data space (Fig.~\ref{figIntensity}). In fact, the cloud defines a hypersurface (manifold) embedded in that space, with its dimensionality determined by the number of degrees of freedom available to the system. Snapshots from a rotating molecule, for example, form a 3D hypersurface.

This perspective naturally leads one to use the tools of differential geometry for data analysis, with the metric playing a particularly important role. In non-technical terms, one would like to relate infinitesimal changes in a given snapshot to the corresponding infinitesimal changes in orientation, giving rise to the changes in the snapshot. In other words, one would like to relate the metric of the data manifold produced by the collection of snapshots to the metric of the manifold of rotations. This would allow one to determine the rotation operation connecting any pair of snapshots. The problem, however, is that the metric of scattering manifolds is not simply related to that of the rotation operations. 
We solve this problem in two steps. First, we show that the metric of the data manifold can be decomposed into two parts, one with high symmetry, and an object-specific ``residual'' with low symmetry. Second, using results from general relativity and quantum mechanics, we show that the eigenfunctions of the high-symmetry part are directly related to those of the manifold of rotations. This allows one to deduce the snapshot orientations from the high-symmetry part, which is the same for all objects, and use the object-specific part as a fingerprint of each object.

We conclude this section with four observations. First, real datasets necessarily contain a finite number of observations. As such, they must be treated by graph-theoretic means, which tend to the differential geometric limit under appropriate conditions. This issue is addressed below, as needed. Second, we have not distinguished between image and diffraction snapshots. If the dataset consists of the latter, each snapshot, or a reconstructed 3D diffraction volume must be inverted by so-called phasing algorithms \cite{GerchbergSaxton72,Fienup78,OszlanyiSuto04}. As this procedure is well established, we do not address it further. Third, the discussion is restricted to the effect of operations on objects with no symmetry. The case where the object itself has specific symmetries will be treated elsewhere. Finally, the knowledge gained in the course of our analysis is sufficient to navigate from any starting point to any desired destination on the manifold; i.e., given any snapshot, produce any other as required. This is, of course, tantamount to possessing a 3D model of the object. As this approach is somewhat unfamiliar, we provide actual 3D models to demonstrate the power of our approach.

\section{Previous work}
\label{PreviousWork}

In order to place our contribution in context, we present a brief summary of previous work.  As a review is beyond the scope this paper, the summary is necessarily brief, leaving out much excellent work in this general area.

When the snapshots emanate from known object orientations and the signal is adequate, standard tomographic approaches \cite{Natterer01} are routinely employed. 3D models can also be constructed when the snapshot orientations are unknown and the signal is low \cite{Frank02} [signal-to-noise ratio (SNR) $ \sim -5 $ dB], or extremely low \cite{FungEtAl08, LohElser09,SchwanderEtAl10a} ($\sim10^{-2}$ scattered photons per Shannon pixel with Poisson noise and background scattering). Finally, efforts are underway to recover structure, map conformations, and determine dynamics (``3D movies'') from random sightings of non-identical members of heterogeneous ensembles  \cite{SchwanderEtAl10a,ScheresEtAl07,FischerEtAl10} and/or evolving systems \cite{FischerEtAl10,SchwanderEtAl10a}. 

\begin{figure}
\includegraphics[width=\linewidth]{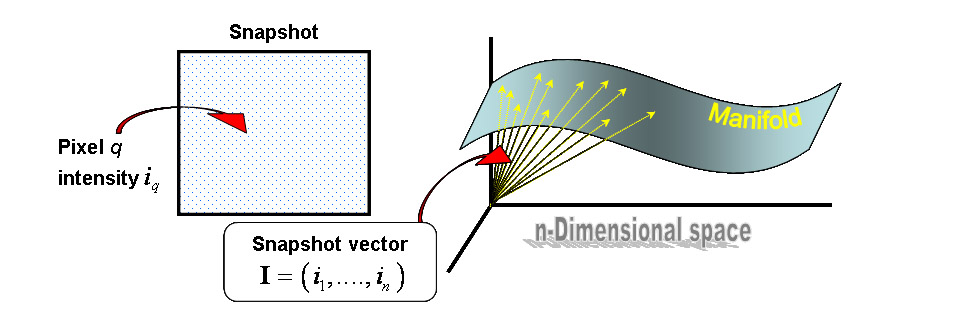}%
\caption{\label{figIntensity}Intensity values of a snapshot on a detector with $ n $ pixels, represented as an $ n $-dimensional vector. Changes in the object or its position alter the snapshot, causing the vector to trace out a manifold in the $ n $-dimensional data space. The dimensionality of manifold is determined by the number of degrees of freedom available to the object. For instance, the rotations of a rigid object give rise to a 3D manifold.}%
\end{figure}

Data-analytical approaches now include powerful graph-theoretic and/or differential geometric means to deduce information from the global structure of the data representing the (nonlinear) correlations in the dataset \cite{TenenbaumEtAl00,RoweisSaul00,BelkinNiyogi03,DonohoEtAl03,CoifmanEtAl05,CoifmanLafon06,YounesEtAl08,LinEtAl06}. Ideally, this structure takes the form of a low-dimensional manifold in some high-dimensional space dictated by the measurement apparatus (see Fig.~\ref{figIntensity}). These so-called manifold-embedding approaches are in essence nonlinear (kernel) principal components techniques \cite{SchoelkopfEtAl98}, which seek to display in some low-dimensional Euclidean space the manifold representing the correlations in the data. While powerful, such approaches face three challenges: computational cost; robustness against noise; and the assignment of physical meaning to the outcome of the analysis, i.e., physically correct interpretation.

Bayesian manifold approaches \cite{FungEtAl08,BishopEtAl98} and their equivalents \cite{LohElser09,MothsOurmazd11} are able to operate at extremely low signal, but require prior knowledge of the manifold dimensionality and its physical meaning.  They also display extremely unfavorable scaling behaviors \cite{FungEtAl08,LohElser09,SchwanderEtAl10a,MothsOurmazd11}.  It has thus not been possible, for example, to reconstruct objects with diameters exceeding eight times the spatial resolution, severely limiting the size of amenable objects and/or the resolution of the reconstruction.

Non-Bayesian graph-theoretic methods are computationally efficient, but tend not to be robust against noise \cite{BalasubramanianSchwartz02,CoifmanEtAl08}. More fundamentally, it has proved difficult to assign physical meaning to the outcome of graph-theoretic analyses \cite{CoifmanEtAl10}, in the sense that the meaning of the dimensions in which the data manifold is embedded is unknown. Strategies to overcome this problem have included exploring the graph-theoretic consequences of specific changes in model systems \cite{FergusonEtAl10}, making restrictive assumptions about the nature of the data, and/or extracting specific information from the data first and subjecting this information to graph-theoretic analysis \cite{CoifmanEtAl10,SingerEtAl10}.  

In this paper, we present a computationally efficient theoretical framework capable of interpreting the outcome of graph-theoretic analysis of scattering data without restrictive assumptions. Using this approach, we demonstrate 3D structure recovery from 2D diffraction snapshots of unknown orientation at computational complexities four orders of magnitude higher than hitherto possible \cite{FungEtAl08, LohElser09}. In Paper~II, we show that this framework can be used to: (1) recover structure from simulated and experimental snapshots at signal levels $ \sim 10 \times $ lower than currently in use; and (2) reconstruct time-series (movies) from ultralow-signal random sightings of evolving objects. In sum, therefore, our approach offers a powerful route to recovering structure and dynamics (3D movies) from ultralow-signal snapshots with no orientational or timing information.

\section{Symmetry-based analysis of scattering data}
\label{secTheory}
 
In this section we develop a mathematical framework for analyzing ensembles of 2D snapshots, using far-field scattering by a single object as a model problem to focus the discussion (see Sec.~\ref{secSetting}). The basic principle of our approach, laid out in Sec.~\ref{secRiemannianFormulation}, is that the data acquisition process can be described as a manifold embedding $ \Phi $ \citep[][]{Schutz80,Lang02}, mapping the set of orientations of the object, SO(3), to the Hilbert space of snapshots on the detector plane. As a result, the differential-geometric properties of the rotation group formally carry over to the scattering dataset. In particular, the dataset can be equipped with a homogeneous Riemannian metric $ B $, whose Laplacian eigenfunctions are the well-known Wigner $ D $-functions \cite{Wigner59,BiedenharnLouck81,ChirikjianKyatkin00}. Here, a metric is called homogeneous if any two points on the data manifold can be connected via a transformation that leaves the metric invariant. We refer to this class of transformations as symmetries. In Sec.~\ref{secSymmetries}, we show that taking advantage of symmetry, as manifested in the properties of homogeneous metrics on SO(3) \citep{Hu73}, leads to a powerful means for recovering snapshot orientations and hence the 3D structure of objects.   

A number of sparse algorithms  \cite{BelkinNiyogi03,CoifmanLafon06} are able to compute discrete approximations of Laplacian eigenfunctions directly from the data. However, these algorithms do not provide the eigenfunctions associated with $ B $, but rather the eigenfunctions of an induced metric $ g $ associated with the embedding $ \Phi $ (i.e., the measurement process). The properties of this induced metric are discussed in Sec.~\ref{secInducedMetric}. There, we show that $ g $ is not homogeneous, but admits a decomposition into a homogeneous metric plus a residual. Intriguingly, the homogeneous part of $ g $ corresponds to a well-known solution of general relativity (the so-called Taub solution \cite{Taub51}), which has the important property of preserving the Wigner $ D $-functions as solutions of the Laplacian eigenvalue problem \cite{Hu73}. As described in Sec.~\ref{secExtension} (and demonstrated in Sec.~\ref{secResults} and Paper~II), this property applies to a broad range of scattering modalities, and can be exploited to perform highly accurate 3D reconstruction in a computationally-efficient manner.   

\subsection{Image formation}
\label{secSetting}
 
\begin{figure}
\includegraphics[width=\linewidth]{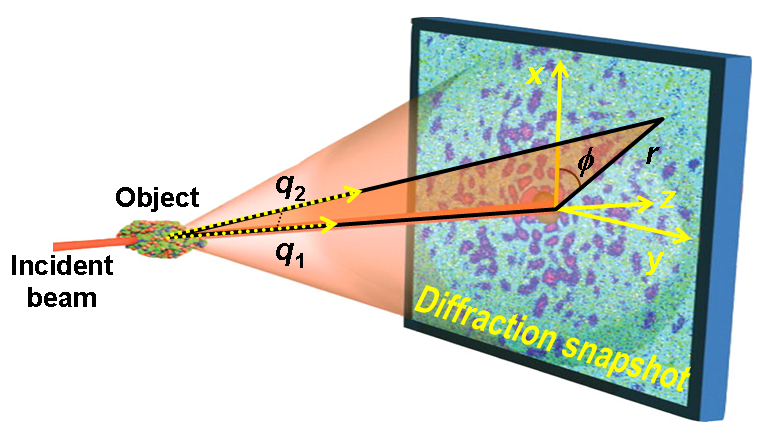}
  \caption{\label{figGeometry}Geometry of image formation by scattering. $ \bm{ q }_1 $ and $ \bm{ q }_2 $ represent the wavevectors for the incident and scattered radiation, respectively, $ ( x, y, z ) $ the lab frame, and $ ( r, \phi ) $ coordinates in the detector plane.}
\end{figure}

We first treat image formation as elastic, kinematic scattering of unpolarized radiation onto a far-field detector in reciprocal space \cite{Cowley95}, where each incident photon can scatter from the object at most once (kinematic scattering), and energy is conserved (elastic scattering).  We show later that our conclusions are more generally applicable. In this minimal model, illustrated in Fig.~\ref{figGeometry}, an incident beam of radiation with wavevector $ \bm{ q }_1 = Q \bm{ z } $ (we set $ Q > 0 $ by convention) is scattered by an object of density $ \rho( \bm{ u } ) $ with Patterson function $ P( \bm{ u } ) = \int d \bm{ u }' \, \rho( \bm{ u }' ) \rho( \bm{ u } - \bm{ u }' )  = P( -\bm{ u } ) $, where $ \bm{ u } $ and $ \bm{ u }' $ are position vectors in $ \mathbb { R }^ 3 $. The structure-factor amplitude at point $ \vec{ r } $ on a detector plane fixed at right angles to the incident beam is given by the usual integral, 
\begin{equation}
  \label{eqDiffractionIntensity}
  a( \vec{ r } ) = \left[ \omega( \vec{ r } ) \int d\bm{ u } \, P( \bm{ u } ) \exp\bm{ ( } i \bm{ u }  \bm{ \cdot } \bm{ q }( \vec{ r } ) \bm{ ) } \right]^{1/2},
\end{equation} 
where $ \omega( \vec{ r } ) $ is an obliquity factor proportional to the solid angle subtended at $ \bm{ u } = \bm{ 0 } $  by the detector element at $ \vec{ r } $, and $ \bm{ q }( \vec{ r } ) $ is the change in wavevector due to scattering. In elastic scattering we have
\begin{equation}
  \label{eqScatteredQ}
  \begin{gathered}
    \bm{ q }( r ) = \bm{ q }_2( \vec{ r } ) - \bm{ q }_1, \\
    \bm{ q }_2( \vec{ r } ) = Q( \sin\theta \cos\phi, \sin\theta \sin\phi, \cos\phi ), 
  \end{gathered}
\end{equation}
where $ \bm{ q }_2( \vec{ r } ) $ is the scattered wavevector, and $ \theta $ and $ \phi $ are the polar and azimuthal angles in reciprocal space corresponding to position $ \vec{ r } $ on the detector plane (see Fig.~\ref{figGeometry}). Note that, by the convolution theorem, the Fourier transform $ \mathcal{ F }( P ) $ of the Patterson function is equal to the non-negative function $ | \mathcal{ F }( \rho ) |^2 $. As a result, no modulus sign is needed in Eq.~\eqref{eqDiffractionIntensity}, and $ a^2( \vec{ r }_j ) $ is equal to the intensity $ I_j $ at pixel $ j $ in Fig.~\ref{figIntensity}. 

In an idealized, noise-free experiment involving a single object conformation, one observes a sequence of $ s $ snapshots on a detector of infinite extent, with the snapshots arising from random orientations of the object. Each snapshot is obtained from Eq.~\eqref{eqDiffractionIntensity} by replacing $ P( \bm{ u } ) $ with $ P_\mathsf{ R }( \bm{ u } ) = P( \mathsf{ R }^{-1} \bm{ u } ) $, where $ \mathsf{ R } $ is a $ 3 \times 3 $ right-handed rotation matrix; i.e., $ \mathsf{ R }^\text{T} \mathsf{ R } = \mathsf{ I } $ and $\det( \mathsf{ R } ) = 1 $. 

The set of all matrices satisfying these conditions form the 3D rotation group, SO(3). It is well known that SO(3) is a Lie group, i.e., it is a differentiable manifold (in this case of dimension~3) \citep{Schutz80,ChirikjianKyatkin00,Arvanitoyeorgos03}. Among the several parameterizations (coordinate charts) of SO(3), of interest to us here will be Euler angles, unit quaternions, and hyperspherical coordinates \cite{Kuipers02}. The last coordinate system stems from the fact that SO(3) has the topology of a three-sphere with its antipodal points identified. 

For the remainder of this section, SO(3) will play the role of the \emph{latent manifold} $ \mathcal{ S } $, i.e., the set of degrees of freedom available to the object. In more general applications, the latent manifold would be augmented to contain the additional degrees of freedom, provided, of course, that these degrees of freedom admit a manifold description---a natural requirement for operations such as shifts and smooth conformational changes.      

\subsection{Riemannian formulation of image formation}
\label{secRiemannianFormulation}

We are concerned with intensity patterns generated by scattering, i.e., a subset of all possible patterns on a planar detector, referred to as \emph{data space}. It is reasonable to expect that in a physical experiment involving a finite-power beam, the resulting distributions of structure-factor moduli belong to the set of square-integrable functions on the detector plane, $ L^2( \mathbb{ R }^2 ) $. This is a Hilbert space of scalar functions $ f_i $ equipped with the usual inner product  
\begin{equation}
  \label{eqL2InnerProduct}
  ( f_1, f_2 ) = \int d \vec{ r } \, f_1( \vec{ r } ) f_2( \vec{ r } ),
\end{equation} 
and corresponding norm 
\begin{equation}
  \label{eqL2Norm}
  \lVert f_i \rVert = ( f_i, f_i )^{1/2}.
\end{equation}
In many respects, $ L^2( \mathbb{ R }^2 ) $ can be though of as a generalization of Euclidean space to infinite dimensions. In particular, the $ L^2 $ norm induces a distance $ \lVert f_1 - f_2 \rVert $ analogous to the standard distance in finite-dimensional Euclidean space. Moreover, viewed as a manifold \cite{Lang98}, $ L^2( \mathbb{ R }^2 ) $ has the important property that its elements are in one-to-one correspondence with its tangent spaces. Thus, the inner product in Eq.~\eqref{eqL2InnerProduct} can be interpreted as a metric tensor acting on pairs of tangent vectors on $ L^2( \mathbb{ R }^2 ) $, or manifolds embedded in $ L^2( \mathbb{ R }^2 ) $.

We describe the image formation process as an embedding \cite{Schutz80,Lang02}, $ \Phi : \mathcal{ S } \mapsto L^2( \mathbb{ R }^2 ) $, taking the latent manifold into data space. For instance, in the present application with $ \mathcal{ S } = \text{SO(3)} $, we have the explicit formula $ \Phi( \mathsf{ R } ) = a_\mathsf{ R } $, where $ \mathsf{ R } $ is an SO(3) rotation matrix and $ a_\mathsf{ R } $ is the pattern on the detector given by Eq.~\eqref{eqDiffractionIntensity} with $ P( \bm{ u } ) $ replaced by $ P( \mathsf{ R }^{-1} \bm{ u } ) $. 

The image $ \mathcal{ M } = \Phi( \mathcal{ S } ) $ of the latent manifold in data space is called the \emph{data manifold} (see Fig.~\ref{figIntensity}). In the absence of degeneracies such as object symmetry, the data and latent manifolds are diffeomorphic manifolds, i.e., completely equivalent from the point of view of differential geometry.  In contrast with manifolds of shapes \cite{BronsteinEtAl07,YounesEtAl08} which can contain singularities, these are manifolds of operations and thus generally well-behaved. The image of the latent manifold in data space is then a smooth (here, three-dimensional) embedded hypersurface, which preserves the topology of the latent manifold, and the structure of its tangent spaces \cite{SauerEtAl91}. Perception, at least in this simple model, can be viewed as understanding the map between the latent manifold of orientations and the data manifold of pixel intensities. 

The inner product induced on the data manifold by the inner product in data space leads to a metric tensor $ g $ on the latent manifold, which encodes the properties of the object and the imaging process. This metric is is constructed by converting the inner product of $ L^2( \mathbb{ R }^2 ) $ in Eq.~\eqref{eqL2InnerProduct} to an equivalent inner product between tangent vectors on the data manifold. Specifically, given tangent vectors $ v_1 $ and $ v_2 $ on $ \mathcal{ S } $, the induced metric $ g $ is defined through the action
\begin{equation}
  \label{eqGInduced}
  g( v_1, v_2 ) = \bm{ ( } \Phi^*( v_1 ), \Phi^*( v_2 ) \bm{ ) },
\end{equation}
where $ \Phi^* $ is the so-called derivative map associated with $ \Phi $ \cite{Wald84,Lang98}, carrying along tangent  vectors on $ \mathcal{ S } $ to tangent vectors on $ \mathcal{ M } $. The induced metric can be expanded in a suitable tensorial basis for $ \mathcal{ S } $ as a $ 3 \times 3 $ symmetric positive-definite (SPD) matrix with components $ g_{\mu\nu} $, viz.,
\begin{equation}
  \label{eqGExpansionR}
  g = \sum_{\mu,\nu=1}^3  g_{\mu\nu} E^\mu E^\nu,
\end{equation}
where $ \{ E^1, E^2, E^3 \} $ is a basis of dual vector fields on $ \mathcal{ S } $. Note that $ ds^2 = g( \delta v, \delta v ) $ corresponds to the squared distance between two nearby points on $ \mathcal{ S } $ with relative separation $ \delta v $. The integral of $ ds^2 $ over curves on $ \mathcal{ S } $ provides the data manifold with a notion of length and distance.   

For the purposes of the symmetry analysis ahead, it is useful to consider expansions of $ g $ in a basis of right-invariant vector fields \cite{Schutz80,Arvanitoyeorgos03}, where, following the derivation in Appendix~\ref{appInducedMetric}, the components of $ g $ at orientation $ \mathsf{ R } $ are found to be
\begin{equation}
  \label{eqGComponentsR}
  g_{\mu\nu}( \mathsf{ R } ) = -\int d\bm{ r } \, \omega( \vec{ r } ) \left. \bm{ \nabla } \bm{ \cdot } [\mathsf{ J }_\mu \bm{ q } a_\mathsf{ R }( \bm{ q } ) ] \bm{ \nabla } \bm{ \cdot } [ \mathsf{ J }_\nu \bm{ q } a_\mathsf{ R }( \bm{ q } ) ] \right\rvert_{ \bm{ q }( \vec{ r } ) }.
\end{equation}
In the above, $ \int d\bm{ r } $ denotes integration over the detector plane; $ \bm{ q }( \vec{ r } )  $  is the change in wavevector given by Eq.~\eqref{eqScatteredQ}; $ \bm{ \nabla } = ( \partial / \partial q_x, \partial / \partial q_y, \partial / \partial q_z ) $ is the gradient operator in reciprocal space; and $ \mathsf{ J }_\mu $ are the $ 3 \times 3 $ antisymmetric matrices in Eq.~\eqref{eqJMu} generating rotations about the $ x $, $ y $, and $ z $ axes, respectively. As stated above, $ a_\mathsf{ R } $ is the structure-factor amplitude corresponding to orientation $ \mathsf{ R } $.

\subsection{\label{secSymmetries}Symmetries}

We now show that the Riemannian formulation described above reveals important symmetries, which can be used to determine the object orientation separately from the object itself. A fundamental concept in the discussion below is the notion of an \emph{isometry} \cite{Schutz80,Wald84}. Broadly speaking, an isometry is a continuous invertible transformation that leaves $ g $ invariant. More specifically, any diffeomorphism $ \phi : \mathcal{ S } \mapsto \mathcal{ S } $ mapping the latent manifold to itself induces a transformation $ \phi^* $ acting on tensors of the manifold; the latter will be an isometry if $ \phi^* ( g ) = g $ holds everywhere on $ \mathcal{ S } $. A symmetry, therefore, in this context is an operation that leaves distances on the data manifold unchanged. 

If the group of isometries of $ g $ acts transitively on $ \mathcal{ S } $ (i.e., any two points in $ \mathcal{ S } $ can be connected via a $ \phi $ transformation), the pair $  ( \mathcal{ S }, g ) $ becomes a \emph{Riemannian homogeneous space} \cite{Arvanitoyeorgos03}. We refer to any $ g $ meeting this condition as a homogeneous metric. Riemannian homogeneous spaces possess natural sets of orthonormal basis functions, analogous to the Fourier functions on the line and the spherical harmonics on the sphere, which can be employed for efficient data analysis \cite{ChirikjianKyatkin00}. It is therefore reasonable to design algorithms that explicitly take into account the underlying Riemannian symmetries of scattering data sets.  

As discussed in more detail below, the isometry group of $ g $ in Eq.~\eqref{eqGInduced} is generally not large-enough to induce a transitive action; i.e., there exist points on the manifold that cannot be mapped to one another through an isometry. Nevertheless, considerable progress in the interpretation of datasets produced by scattering can be made by establishing the existence of a related metric $ h $, whose isometry group meets the conditions for transitivity. As shown below, the properties of $ h $ can be used to interpret the results of data analysis performed on $( \mathcal{ S }, g ) $. By combining aspects of group theory and differential geometry, the general techniques developed here constitute a novel approach for analyzing scattering data, and potentially other machine-learning applications.            

The canonical classes of symmetry operations in problems involving orientational degrees of freedom are the so-called left and right multiplication maps \cite{Schutz80,Wald84,Arvanitoyeorgos03}. These maps, respectively denoted $ L_\mathsf{ Q } $ and $ R_\mathsf{ Q } $, are parameterized by an arbitrary rotation matrix $ \mathsf{ Q } $ in $ \text{SO(3)} $, and act on SO(3) elements by multiplication from the left or the right, respectively. That is, 
\begin{equation}
  \label{eqLeftRightMultiplication}
  L_\mathsf{ \mathsf{ Q } }( \mathsf{ R } ) = \mathsf{ Q } \mathsf{ R }, \quad R_\mathsf{ Q }( \mathsf{ R } ) = \mathsf{ R } \mathsf{ Q }.
\end{equation}
Viewed as groups, the collection of all left multiplication maps or right multiplication maps are isomorphic to SO(3), from which it immediately follows that their action is transitive. Thus, any metric $ g $ invariant under $ L_\mathsf{ Q } $ or $ R_\mathsf{ Q } $ (or both) makes $ ( \mathcal{ S }, g ) $ a Riemannian homogeneous space. 

The natural metric for the SO(3) latent manifold of orientations with these symmetries is the metric tensor $ B $ associated with the Killing form of the Lie algebra of SO(3) \cite[][]{Schutz80,Arvanitoyeorgos03}. This metric is bi-invariant under left and right translations. The corresponding group of isometries has $ \text{SO(3)} \times \text{SO(3)} $ structure, i.e., it is a six-dimensional group. In hyperspherical coordinates $ ( \chi, \theta, \phi ) $, $ B $ is represented by the diagonal matrix
\begin{equation}
  \label{eqRoundMetric}
  [ B_{\mu\nu} ] = \diag( 1, \sin^2\chi, \sin^2\chi \sin^2\theta ),
\end{equation}
which is identical to the canonical metric on the three-sphere $ S^3 $. For this reason $ B $ is frequently referred to as the ``round'' metric on SO(3).

A key implication of the symmetries of $ B $ pertains to the eigenfunctions of the corresponding Laplace-Beltrami operator $ \upDelta_B $ \cite{Berard86}, defined as
\begin{equation}
  \label{eqLaplB}
  \upDelta_B( f ) = - | B |^{-1/2} \sum_{\mu,\nu=1}^m \partial_\mu \left( | B |^{1/2} B^{\mu\nu} \partial_\nu f  \right)
\end{equation} 
for $ | B | = \det [ B_{\mu\nu} ] $, $ [ B^{\mu\nu} ] = [ B_{\mu\nu}]^{-1} $, and a scalar function $ f $ on SO(3). It is a well-known result in harmonic analysis that the eigenfunctions of $ \upDelta_B $ are  the Wigner $ D $-functions \cite{Vilenkin68,VarshalovichEtAl88,ChirikjianKyatkin00}, complex-valued functions on SO(3), which solve the eigenvalue problem 
\begin{equation}
  \label{eqEigenvalueProblemD}
  \upDelta_B D^j_{mm'}( \mathsf{ R } )= j( j + 1 ) D^j_{mm'}( \mathsf{ R } )
\end{equation} 
with positive integer $ j $ and integers $ m $ and $ m' $ in the range $ [ -j, j ] $. Written out in terms of Euler angles in the $zyz$ convention, $ ( \alpha^1, \alpha^2, \alpha^3 ) $, explicit formulas for the $ j = 1 $ $ D $-functions are
\begin{equation}
  \label{eqD1}
  \begin{aligned}
    D^1_{00}( \alpha^1, \alpha^2, \alpha^3 ) &= \cos \alpha^2, \\
    D^1_{\pm 1 0 }( \alpha^1, \alpha^2, \alpha^3 ) &=  e^{\mp i \alpha^1 } \sin( \alpha^2 ) / 2^{1/2}, \\
    D^1_{0,\pm 1}( \alpha^1, \alpha^2, \alpha^3 ) &= \sin( \alpha^2 ) e^{\pm i \alpha^3} / 2^{1/2}, \\
    D^1_{-11}( \alpha^1, \alpha^2, \alpha^3 ) &= e^{-i( \alpha^1+\alpha^3)} \cos^2( \alpha^2/2), \\
    D^1_{-1-1}( \alpha^1, \alpha^2, \alpha^3) &= e^{i(\alpha^1+\alpha^3)} \cos^2( \alpha^2 / 2 ), \\
    D^1_{-11}( \alpha^1, \alpha^2, \alpha^3 ) &= e^{i(\alpha^1-\alpha^3)} \sin^2( \alpha^2 / 2 ), \\
    D^1_{1-1}( \alpha^1, \alpha^2, \alpha^3 ) &= e^{-i( \alpha^1 - \alpha^3) } \sin^2( \alpha^2/2).
  \end{aligned}
\end{equation}

As one can check by algebraic manipulation, the above ninefold-degenerate set of eigenfunctions can be put into one-to-one correspondence with the elements of $ \mathsf{ R } $. That is, it is possible to find linear combination coefficients $ c_{pqmm'} $ such that 
\begin{equation}
  \label{eqRDCorrespondence}
  R_{pq} = \sum_{m,m'=-1}^1 c_{pqmm'} D^1_{mm'}( \mathsf{ R } ),
\end{equation}
where $ R_{pq} $ are the  elements of $ \mathsf{ R } $. Thus, if eigenfunctions of $ \upDelta_B $ could be accessed by processing the observed snapshots $ a_\mathsf{ R } $ on the data manifold, e.g., through one of the graph-theoretic algorithms developed in machine learning \cite[][]{BelkinNiyogi03,CoifmanLafon06}, Eq.~\eqref{eqRDCorrespondence} could be used to invert the embedding map $ \Phi $. This would be tantamount to having learned to ``navigate'' on the data manifold.

The method outlined above for symmetry-based inversion of $ \Phi $ is also applicable under weaker symmetry conditions on the metric. In particular, it can be extended to certain metrics of the form
\begin{equation}
  \label{eqHuMetric}
  h = \ell_1 E^1 E^1 + \ell_2 E^2 E^2 + \ell_3 E^3 E^3,
\end{equation}
where $ E^\mu $ are the right-invariant dual basis vectors in Eq.~\eqref{eqEDual}, and $ \ell_\mu $ are non-negative parameters. These are the so-called spinning-top metrics of classical and quantum-mechanical rotors \cite{BiedenharnLouck81,McCauley97}, arising also in homogeneous cosmological models of general relativity \cite{Taub51,Misner69,Hu73}. In classical and quantum mechanics, $ h $ features in the Hamiltonian of a rotating rigid body with principal moments of inertia given by $ \ell_\mu $. In general relativity, $ \ell_\mu$ characterize the anisotropy of space in the so-called mixmaster cosmological model \cite{Misner69}. 

By construction, metrics in this family are invariant under arbitrary right multiplications, i.e., they possess an SO(3) isometry group. This is sufficient to make $ ( \mathcal{ S }, h ) $ a Riemannian homogeneous space. If two of the $ \ell_\mu $ parameters are equal (e.g., $ \ell_1 = \ell_2 $), $ h $ describes the motion of an axisymmetric rotor, or the spatial structure of the Taub solution in general relativity \cite{Taub51}. As one may explicitly verify, in addition to being invariant under left multiplications, the metric of the axisymmetric rotor is invariant under translations from the left by the one-parameter subgroup associated with rotations about the axis of symmetry, which corresponds here to the direction of the incoming scattering beam. Therefore, it has a larger, $ \text{SO(3)} \times \text{SO(2)} $ isometry group than the more general metric of an asymmetric rotor. In the special case that all of the $ \ell_\mu $ are equal, $ h $ reduces to a multiple of the round metric $ B $ in Eq.~\eqref{eqRoundMetric}. 

Let $ \upDelta_h $ denote the Laplace-Beltrami operator associated with $ h $. It is possible to verify that $ \upDelta_h $, and the Laplacian $ \upDelta_B $ corresponding to the round metric commute. That is, it is possible to find eigenfunctions that satisfy the eigenvalue problem for $ \upDelta_B $ and $ \upDelta_h $ simultaneously. In particular, as Hu \cite{Hu73} shows, the eigenvalue-eigenfunction pairs $ ( \lambda^j_m, y^j_m ) $ of $ \upDelta_h $ can be expressed as linear combinations of Wigner $ D $-functions with the same $ j $ and $ m $ quantum numbers: 
\begin{equation}
  \label{eqY}
  y^j_m = \sum_{m'=-j}^j A^j_{mm'} D^j_{mm'},
\end{equation}
for some linear expansion coefficients $ A^j_{mm'} $. For the most general metrics with $ \ell_1 \neq \ell_2 \neq \ell_3 $, no closed-form expression is available for either $ A^j_{mm'} $, or the corresponding eigenvalues $ \lambda^j_m $. However, in the special case of axisymmetric rotors, any $ A^j_{mm'} $ in Eq.~\eqref{eqY} will produce an eigenfunction of $ \upDelta_h $. This means that the vector space spanned by the $ j = 1 $ eigenfunctions associated with the metric of an axisymmetric rotor (denoted here by $y^j_{mm'}$) is nine-dimensional. In particular, Eq.~\eqref{eqRDCorrespondence} can be applied directly with $ D^1_{mm'} $ replaced by $ y^1_{mm'} $.

The eigenvalue $ \lambda^j_m $ corresponding to $ y^j_{mm'} $ in the so-called prolate configuration, $ \ell_1 = \ell_2 \leq \ell_3 $, is given by
\begin{equation}
  \label{eqLambdaSymmetricTop}
  \lambda^j_m = \frac{ 1 }{ 2 \ell_1 } j( j + 1) - \left( \frac{ 1 }{ 2 \ell_1 } - \frac{ 1 }{ 2 \ell_3 } \right) m^2,
\end{equation}
with a similar formula for the oblate configuration $ \ell_1 = \ell_2 \geq \ell_3 $ \cite{Hu73}. An important point about Eq.~\eqref{eqLambdaSymmetricTop} (and its oblate analog) is that all of the $ j = 1 $ eigenvalues are greater than zero and smaller than the maximal $ j = 2 $ eigenvalue; i.e., the $ j = 1 $ eigenfunctions can be identified by ordering the eigenvalues. This property, which does not apply for general asymmetric-rotor metrics, alleviates the difficulty of identifying the correct subset of eigenfunctions for orientation recovery via Eq.~\eqref{eqRDCorrespondence}.  
         
\subsection{Accessing the homogeneous metric}
\label{secInducedMetric}

The method outlined above for symmetry-based inversion of $ \Phi $ makes direct use of the fact that $ ( \mathcal{ S }, h ) $ is a Riemannian homogeneous space. However, $ ( \mathcal{ S }, h ) $ is generally inaccessible to numerical algorithms, which access a discrete subset of the latent manifold equipped with the induced metric $ g $, i.e., $ ( \mathcal{ S }, g ) $. It is therefore important to establish the isometries of $ g $, because, for instance, the latter govern the behavior of the Laplacian eigenfunctions computed via graph-theoretic algorithms \cite{TenenbaumEtAl00,RoweisSaul00,CoifmanEtAl05,BelkinNiyogi03,CoifmanLafon06}.

In general, $ g $ is not invariant under arbitrary left or right multiplications. The lack of invariance of $ g $ under the right multiplication map $ R_\mathsf{ Q } $ can be seen by considering its components in a right-invariant basis. In particular, it can be shown that $ R_\mathsf{ Q } $ is an isometry of $ g $ if and only if the components $ g_{\mu\nu} $ in the right-invariant basis from Eq.~\eqref{eqGComponentsR} are invariant under replacing $ \mathsf{ R } $ by $ \mathsf{ R } \mathsf{ Q } $, which does not hold generally. 

What about the behavior of $ g $ under left multiplications? Here, an expansion of $ g $ in an analogous left-invariant basis reveals that $ g $ is not invariant under left multiplications by general SO(3) matrices, but is invariant under left multiplication by a rotation matrix $ \mathsf{ Q } $ about the beam axis $ z $ (see Fig.~\ref{figGeometry}). This is intuitively obvious, because the outcome of replacing a snapshot $ a_{\mathsf{R}} $ with $ a_{\mathsf{Q}\mathsf{R}} $ is then a rotation on the detector plane, which has no influence on the value of the $ L^2 $ inner product used to evaluate $ g $. In summary, instead of the six-dimensional $ \text{SO(3)} \times \text{SO(3)} $ isometry group of $ B $, the isometry group of the induced metric is the one-dimensional SO(2) group of rotations about the beam axis, which obviously cannot act transitively on the three-dimensional latent manifold. 

Despite the insufficiency of $ g $ to make $ ( \mathcal{ S }, g ) $ a Riemannian homogeneous space, $ g $ admits a decomposition of the form
\begin{equation}
  \label{eqGDecomposition}
  g = h + w,
\end{equation}
with the following key properties: (1) $ h $ is a spinning-top metric from Eq.~\eqref{eqHuMetric} with $ \ell_1 = \ell_2 $ (i.e., an axisymmetric-rotor metric, not an arbitrary right-invariant metric). (2) $ w $ is a symmetric (but not necessarily positive-definite) tensor, whose components average  to zero over the manifold, describing the inhomogeneous part of $ g $ (see Appendix~\ref{appDecomposition} for a derivation). Moreover, provided that $ w $ meets a suitable norm constraint, $ g $ and $ h $ are bi-Lipschitz equivalent. This means that the distortion in distances on $ \mathcal{ S } $ caused by using $ g $ instead of $ h $, and the corresponding change in the Laplace-Beltrami eigenfunctions, are both bounded \cite{BerardEtAl94}. The fact that $ h $ in Eq.~\eqref{eqGDecomposition} is the metric of an axisymmetric rotor  is a direct consequence of projection onto a circularly-symmetric 2D detector, and therefore an essential aspect of scattering experiments.  

The important point is this: scattering manifolds are associated with induced metrics $ g $, which, in themselves, possess low symmetry. But they can be decomposed into a homogeneous part $ h $ with a high $ \text{SO(2)} \times \text{SO(3)} $ symmetry and thus associated with well-known Laplacian eigenfunctions independently of the object, plus a low-symmetry residual, which depends on the object. In essence, one can regard the homogeneous metric $ h $ as an object-independent ``Platonic ideal''  \cite{Ross51} version of $ g $, and develop algorithms, which make use of the properties of $ h $ to analyze all scattering datasets.

The symmetries of $ h $ can be exploited in several ways: (1) enforce appropriate constraints in Bayesian algorithms \cite{FungEtAl08}; (2) interpret the results of numerical Ricci flow \cite{CaoZhu06,Topping06}; or (3) approximate the eigenfunctions associated with the homogeneous metric by a truncated set of eigenfunctions associated with the inhomogeneous metric. The last possibility is described below.

Represent each snapshot $ a_{\mathsf{ R }_i } $ in the dataset as a point in nine-dimensional Euclidean space $ \mathbb{ R }^9 $ with coordinates given by $ ( \psi_{i1}, \ldots, \psi_{i9} ) $. Here, $ \psi_{ik} $ are eigenvector components of an elliptic operator on a graph constructed from the $ s $ observed snapshots $ ( a_{\mathsf{ R }_1 }, \ldots, a_{\mathsf{ R }_s } ) $ in $ n $-dimensional data space (see Appendix~\ref{appAlgorithms}). This induces an embedding $ \Psi $ of the latent manifold $ \mathcal{ S } $ in $ \mathbb{ R }^9 $ given by
\begin{equation}
  \label{eqPsi}
  \Psi( \mathsf{ R }_i ) = ( \psi_{i1}, \ldots, \psi_{i9} ).
\end{equation}
Provided that the number of observed snapshots is sufficiently large, suitable algorithms (such as Diffusion Map of Coifman and Lafon \cite{Lafon04,CoifmanLafon06}), lead to embedding coordinates $ \psi_{ik} $, which converge to the corresponding values of the Laplacian eigenfunctions associated with $ g $, even if the sampling of the data manifold is non-uniform. More specifically, for large-enough $ s $ we have the correspondence 
\begin{equation}
  \label{eqPsiApprox}
  \psi_{ik} \approx \psi_k( \mathsf{ R }_i ), 
\end{equation}
where $ \psi_k( \mathsf{ R }_i ) $ is the $ k $-th eigenfunction of the Laplace-Beltrami operator $ \upDelta_g $ associated with the induced metric in Eq.~\eqref{eqGInduced}; i.e., 
\begin{equation}
  \upDelta_g \psi_k = \lambda_k \psi_k
\end{equation}
with
\begin{equation}
  \label{eqLapl}
  \upDelta_g( \cdot ) = - | g |^{-1/2} \sum_{\mu,\nu=1}^m \partial_\mu \left[ | g |^{1/2} g^{\mu\nu} \partial_\nu (\cdot)  \right].
\end{equation}

Now we explicitly employ the symmetries of the homogeneous metric $ h $ in Eq.~\eqref{eqHuMetric}, motivated by the decomposition of $ g $ in Eq.~\eqref{eqGDecomposition}: Express each eigenfunction of the homogeneous Laplacian $ \upDelta_h $ with $ j = 1 $ as a linear combination of the eigenfunctions $ \psi_k $ of the inhomogeneous Laplacian, viz.,
\begin{equation}
   D^1_{mm'}= \sum_{k} c_{mm'k} \psi_k,
\end{equation}
where we have made use of the correspondence in Eq.~\eqref{eqY} between Wigner $ D $-functions and the eigenfunctions associated with the axisymmetric rotor. These eigenfunctions are related to the matrix elements of the rotation operators $ \mathsf{ R }_i $, in accordance with Eq.~\eqref{eqRDCorrespondence}. Retaining only the first nine $ \psi_k $, determine the elements of the rotation matrix by a least-squares fit [Eqs.~\eqref{eqNLSQ1}] of the coordinates computed in Eq.~\eqref{eqPsiApprox} via graph-theoretic analysis. 

The symmetries of $ h $ play an essential role in this procedure, since they yield: (1) the number and sequence of graph-Laplacian eigenfunctions used to represent the latent manifold in Eq.~\eqref{eqPsi}; and (2) the procedure to infer the orientations $ \mathsf{ R }_i $ from the eigenfunctions [in particular, via the structure of the error functional in Eq.~\eqref{eqErrorFunctional}]. As demonstrated in Sec.~\ref{secResults}, this leads to accurate orientation determination for each snapshot.   
 
\subsection{Extension to general scattering problems}
\label{secExtension}

Because the above analysis was performed under a restrictive set of experimental conditions, we now consider the effect of removing these. The assumption of kinematic (single) scattering introduces an additional symmetry due to Friedel's law. As we have not used this symmetry, our arguments remain valid under multiple scattering conditions. The use of linearly or elliptically polarized radiation introduces a second preferred direction in addition to that of the beam, thus removing the SO(2) isometry under left translations, but this can be restored by appropriate correction with the polarization factor. A detector not at right angles to the beam axis also reduces the isometry to $ \text{I} \times \text{SO(3)} $, but this can also be easily corrected by an appropriate geometric factor. Absorption can be accommodated as a complex density function, inelastic scattering by allowing $ \bm{ q }_1 \neq \bm{ q }_2 $, etc. Neither affects our conclusions. Our approach is thus applicable to a wide range of image formation modalities.  

Note, however, that the method has to be modified when the object has discrete or continuous symmetries. This is because the data manifold in this case is not SO(3), but the quotient space $\text{SO(3)} / \Gamma $ where $ \Gamma $ is a subgroup of SO(3) representing the object's symmetries. Among the eigenfunctions of the Laplacian on the SO(3) manifold, only those that are constant on $ \Gamma $ ``survive'' in the $ \text{SO(3)}/\Gamma $ environment. Thus, the orientation recovery procedure must be modified depending on the form of the available eigenfunctions. As mentioned in Sec.~\ref{secConceptualOutline}, this issue will be addressed elsewhere. 
     
\section{Demonstration of structure recovery in 3D }
\label{secResults}
It has long been known that the use of problem-specific constraints can substantially increase computational efficiency \cite{LeCunEtAl90}.  By combining wide applicability with class specificity, symmetries represent a particularly powerful example of such constraints. Exploiting these, we here demonstrate successful orientation recovery for a system computationally $ 10^4\times$ more complex than previously attempted \cite{FungEtAl08,LohElser09}. In Paper~II we apply our framework to snapshots of various kinds with extremely low signal. 

\subsection{3D reconstruction from diffraction snapshots with high computational complexity}

We simulated X-ray snapshots of the closed conformation of \emph{E.~coli} adenylate kinase (ADK; PDB descriptor 1ANK) in different orientations to a spatial resolution $ d = 2.45 \text{ \AA} $, using 1 \text{\AA} photons. In this calculation, Cromer-Mann atomic scattering factors were used for the 1656 non-hydrogen atoms \cite{CromerMann68}, neglecting the hydrogen atoms. We discretized the diffraction patterns on a uniform grid of $ n = 126 \times 126 = 15,876 $ detector pixels of appropriate (Shannon) size, taking the corresponding orientations uniformly on SO(3) according to the algorithm in Ref.~\cite{LovisoloDaSilva01}. The number of diffraction patterns required to sample SO(3) adequately is given by $ 8 \pi^2 ( D / d )^3 \approx 8.5 \times 10^5$, where $ D = 54 \text{ \AA} $ is the diameter of the molecule \cite{FungEtAl08}. In our simulation, however, a larger $ D = 72 \text{ \AA} $ diameter was assumed, allowing, e.g., for the possibility to reconstruct the structure of ADK's open conformation. Hence, a total of $ s = 2 \times 10^6 $ patterns were used in the present analysis. 
 
The diffraction patterns were provided to our Diffusion Map-based algorithm with no orientational information, and the orientation of each diffraction pattern was determined by means of the algorithm in Table~\ref{algNoiseFree} \footnote{The algorithm in Table~\ref{algNoiseFree} was executed with the following parameters: number of nearest neighbors in the sparse distance matrix  $ d = 220 $; Gaussian kernel bandwidth $ \epsilon = 1 \times 10^4 $; number of datapoints for least-squares fitting $ r = 8 \times 10^4 $.}. To estimate the difference between the deduced and true orientations, respectively represented by unit quaternions $ \tilde \tau_i $  and $ \tau_i $ [see Eq.~\eqref{eqUnitQuaternion}], we used the RMS internal angular distance error,
\begin{equation}
  \label{eqDelta}
  \varepsilon = \left[ \frac{ 1 }{ s ( s - 1 ) } \sum_{i\neq j}( \tilde D_{ij} - D_{ij} )^2 \right]^{1/2}.
\end{equation}
In the above, $ D_{ij} = 2 \arccos( | \tau_i \bm{ \cdot } \tau_j | )  $ and $ \tilde D_{ij} = 2 \arccos( | \tilde \tau_i \bm{ \cdot }  \tilde \tau_j | ) $ are respectively the true and estimated internal distances between orientations $ \mathsf{ R }_i $ and $ \mathsf{ R }_j $, and $ \bm{ \cdot } $ is the inner product between quaternions. The resulting internal angular distance error in the present calculation was 0.8 Shannon angles. 

\begin{figure}[tb]
  \centering
 \includegraphics[width=\linewidth]{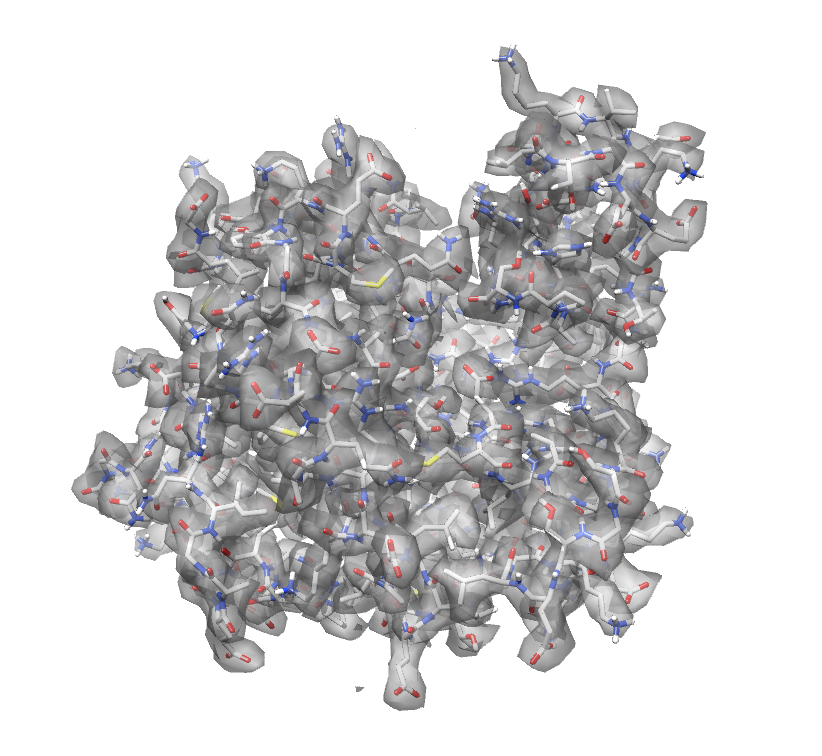}
  \caption{\label{figADK}Three-dimensional electron density of the closed conformation of \emph{E.~coli} adenylate kinase, recovered from $ 2 \times 10^6 $ diffraction patterns of unknown orientation at 2.45 \text{\AA} resolution. Hydrogen atoms were neglected in the electron density calculation. The ball-and-stick model represents the actual structure. See also the EPAPS movie.}
\end{figure}

Next, we placed the diffracted intensities onto a uniform 3D Cartesian grid by ``cone-gridding'' \cite{SchwanderEtAl10b}, and deduced the 3D electron density by iterative phasing with a variant of the charge flipping technique \cite{OszlanyiSuto04} developed by Marchesini \cite{Marchesini08}. Charge flipping was prevented outside a spherical support about twice the diameter of the molecule. The $ R $-factor between the gridded scattering amplitudes and the ones obtained from phasing was $0.19$. The close agreement with the correct structural model is shown in Fig.~\ref{figADK} and the EPAPS movie.

\subsection{Computational cost}
\label{secComputationalCost}

In the absence of orientational information, the computational cost $ C $ of orientation recovery without restrictive sparsity assumptions scales as a power law 
\begin{equation}
  \label{eqR8}
  C \propto R^8 = ( D / d )^8,
\end{equation}
with $ D $ the object diameter and $ d $ the spatial resolution \cite{FungEtAl08,LohElser09,SchwanderEtAl10a}. The analysis of ADK was performed with $ R = 30 $, compared with the largest previously published value of $ R \le 8 $ \cite{FungEtAl08,LohElser09,SchwanderEtAl10a}. This represents an increase of four orders of magnitude in computational complexity over the state of the art, as shown below.  

The computational cost of a single expectation-maximization (EM) step in Bayesian algorithms (e.g., the GTM algorithm \cite{BishopEtAl98,Svensen98}) scales as $ K s n $, where $ K $, $ s $, and $ n $ are the number of quaternion nodes, the number of snapshots, and the data-space dimension, respectively \cite{FungEtAl08}. Here, the data space dimension is equal to the number of Shannon pixels. Assuming an oversampling of 2 in each linear dimension, we have
\begin{equation}
  \begin{aligned}
    K &= \frac{ 8 \pi^2 }{ S } \left( \frac{ D }{ d } \right)^3, \\
    s & = f K = \frac{ 8 \pi^2 f}{ S } \left( \frac{ D }{ d } \right)^3, \\
    n &= \left\{ \frac{ 4 D }{ \lambda } \tan \left[ 2 \arcsin \left( \frac{ \lambda }{ 2 d } \right) \right] \right\}^2 \approx 16 \left( \frac{ D }{ d } \right)^2,  
  \end{aligned}
\end{equation}
where $ S $ is the object symmetry, $ f $ the number of snapshots per orientational bin, and $ \lambda $ the wavelength. Assuming that the number of EM iterations is constant, \citet[][]{FungEtAl08} estimate an eighth-power scaling of the form in Eq.~\eqref{eqR8} for the total computational cost of orientation recovery with GTM. The fact that GTM is NP-hard remains moot.

\citet[][]{LohElser09} argue that their EM-based approach scales as
\begin{equation}
  \label{eqLoh}
  C \propto M_\text{rot} \, s \, N_\text{photons},
\end{equation}
where $ M_\text{rot} $ is the number of rotation samples (orientational bins), and that this leads to an $ R^6 $ scaling. This is based on three assumptions: (1) A sparse formulation can be used to replace the number of detector pixels $ n $ with a much smaller number of scattered photons $ N_\text{photons} $. (2) $ N_\text{photons} $ does not depend on $ R $, and thus can be ignored in the scaling behavior. (3) The number of EM iterations is independent of $ R $ (as in Ref.~\cite{FungEtAl08}), and of $ N_\text{photons} $. Note that $ M_\text{rot} $ and the number of GTM bins, $ K $, are equal. 

\begingroup 
\squeezetable
\begin{table}
  \caption{\label{tableCost}Computational complexity of GroEL and ADK}
  \begin{ruledtabular}
    \begin{tabular}{lccc}
      & GroEL & \multicolumn{2}{c}{ADK} \\
      \cline{3-4}
      & & Sparse & No sparse \\
      & & representation & representation \\
      $ N_\text{photons} $ & $ 10^2 $ & $ \sim 10^3 $ & $ 1.6 \times 10^4 $ \\
      $ s $ & $ 10^6 $ & $ 2 \times 10^6 $ & $ 2 \times 10^6 $ \\
      $ M_\text{rot} $ & $ 2.5 \times 10^4 $ & $ 2 \times 10^6 $ & $ 2 \times 10^6 $ \\
      $ M_\text{rot} \,s \, N_\text{photons} $ & $ 2.5 \times 10^{12} $ & $ 4 \times 10^{15} $ & $ 6.4 \times 10^{16} $ \\
      \textbf{Ratio to GroEL} & \textbf{1} & {$\boldsymbol{ 1.6 \times 10^3 }$} & $ \boldsymbol{2.6 \times 10^4} $ \\
    \end{tabular}
  \end{ruledtabular}
\end{table}
\endgroup

Assumption~(1) holds only for very small photon counts and in the absence of significant background scattering. Assumption~(2) is not justified \cite{ShneersonEtAl08}. For a globular protein, the total number of photons scattered to large angle scales linearly with the number of non-H atoms, and hence object volume, i.e., as $ D^3 $. Thus,
\begin{equation}
  \label{eqLoh2}
  \begin{aligned}    
    M_\text{rot} &= \frac{ 8 \pi^2}{ S } \left( \frac{ D }{ d } \right)^3, \\
    s  &= \frac{ 8 \pi^2 f }{ S } \left( \frac{ D }{ d } \right)^3, \\
    N_\text{photons} \sim n &= 16 \left( \frac{ D }{ d } \right)^2,
  \end{aligned}
\end{equation}
leading to the same $ R^8 $ scaling behavior as Fung \emph{et al.} The validity of assumption~(3) remains moot.

Equipped with the estimates from Eqs.~\eqref{eqR8}--\eqref{eqLoh2}, we now compare the computational complexity of our ADK calculation with the reconstructions of the chignolin and GroEL molecules by \citet[][]{FungEtAl08} and \citet[][]{LohElser09}, respectively. Using Eq.~\eqref{eqR8}, the increase in computation complexity on going from chignolin ($ D_\text{chig} = 16 \text{ \AA} $, $ d_\text{chig} = 1.8 \text{ \AA} $), to ADK ($ D_\text{ADK} = 54 \text{ \AA} $, $ D_\text{ADK (support)} = 72 \text{ \AA} $, $ d_\text{ADK} = 2.45 \text{ \AA} $) is
\begin{equation}
  \left( \frac{ D_\text{ADK (support)} / d_\text{ADK} }{ D_\text{chig} / d_\text{chig} } \right)^8 = \left( \frac{ 29.4 }{ 8.9 } \right)^8 = 1.4 \times 10^4. 
\end{equation}
Using the scaling expression of Eq.~\eqref{eqLoh}, the increase in complexity on going from GroEL to ADK ($72 \text{ \AA}$ support) for the \citeauthor{LohElser09} algorithm varies between $ \Ord( 10^3 ) $ and $ \Ord( 10^4 ) $ for sparse and non-sparse representations, respectively (see Table~\ref{tableCost}). As a sparse representation is not generally possible and we have not had to resort to it, the appropriate comparison is $ 10^4 $. 

\section{Conclusions}
\label{secConclusions}
We have shown that a Riemannian formulation of scattering reveals underlying, object-independent symmetries, which stem from the nature of operations in 3D space and projection onto a 2D detector.  These symmetries lead to immediate identification of the natural eigenfunctions of manifolds produced by data from a wide range of scattering scenarios, and thus to physically-based interpretation of the outcome of graph-theoretic analysis of such data.  In practical terms, the ability to access the homogeneous metric offers a computationally efficient route to determining object orientation without object recognition, while the object-dependent term provides a concise fingerprint of the object for recognition purposes. There are tantalizing indications that face perception in higher primates may occur in this way \cite{FreiwaldTsao10}. The ability to use symmetries to navigate on the data manifold is tantamount to efficient machine learning in 3D, in the sense that given any 2D projection, any other can be reconstructed. 

As shown in Paper~II, the manifold itself offers a powerful route to image reconstruction at extremely low signal, because snapshots reconstructed from the manifold achieve higher signal-to-noise ratios than can be obtained by traditional methods relying on classification and averaging. Combined with the ability to sort random snapshots of an evolving system into a time-series, also demonstrated in Paper~II, our approach offers a radically new way for studying dynamic systems in 4D. Fundamentally, the homogeneous metric describes the transformations of objects without reference to any specific object. This is reminiscent of a Platonic Form, from which specific objects emanate \cite{Ross51}. It is therefore tempting to regard the homogeneous manifold as a Platonic Form, from which our perception of three-dimensional objects stems \cite{Heisenberg81}.

\begin{acknowledgments}
We acknowledge valuable discussions with Y.\ LeCun, B.\ Moths, G.\ N.\ Phillips, Jr., R.\ Rosner, W.\ Schr\"oter, and members of the UWM Physics Department. We are indebted to C.\ H.\ Yoon for stimulating discussions and comments on an earlier version of the manuscript, and grateful to W.\ Schr\"oter for pointing out that a link between symmetries and Platonic Forms was also discussed by Werner Heisenberg and Carl Friedrich von Weizs\"acker in the context of elementary particles.  One of us (DG) is grateful to the Random Shapes Program held in 2007 at the Institute for Pure \& Applied Mathematics.  This work was partially supported by the U.S. Department of Energy Office of Science (SC-22, BES) award \#DE-SC0002164 and a UWM Research Growth Initiative award.
\end{acknowledgments}

\appendix

\section{The induced metric tensor of scattering data sets}
\label{appInducedMetric}

Here, we discuss the properties of the induced metric tensor $ g $ in Eq.~\eqref{eqGInduced}. We begin in Appendix~\ref{appDerivation} with a derivation of the explicit form for the components of $ g $ in the right-invariant basis from Eq.~\eqref{eqGComponentsR}. In Appendix~\ref{appDecomposition} we perform the decomposition of $ g $ in Eq.~\eqref{eqGDecomposition} into axisymmetric and inhomogeneous parts. In Appendix~\ref{appErrorBounds} we provide estimates of the error incurred in using the Laplacian eigenfunctions associated with the inhomogeneous metric $ g $ for orientation recovery.  

\subsection{Derivation}
\label{appDerivation}

Following Sec.~\ref{secRiemannianFormulation}, we describe scattering as an embedding $ \Phi $ taking the latent manifold $ \mathcal{ S } $ to the set of square-integrable intensity patterns $ L^2( \mathbb{ R }^2 ) $ on a 2D dimensional detector. Broadly speaking, the induced metric describes an inner product between tangent vectors on the latent manifold $ \mathcal{ S } $ associated with that embedding. More specifically, in accordance with Eq.~\eqref{eqGInduced}, that inner product is computed by ``pushing forward'' tangent vectors on $ \mathcal{ S } $ to manifest space, and applying the canonical Hilbert space inner product in Eq.~\eqref{eqL2InnerProduct}. The map carrying along the tangent vectors of $ \mathcal{ S } $ is the derivative map $ \Phi^* $ associated with $ \Phi $, which is evaluated as follows.  

First, note that every smooth tangent vector field $ v $ on $ \mathcal{ S } $ generates a corresponding one-parameter family of transformations \cite{Schutz80,Lang02,Arvanitoyeorgos03}, 
\begin{equation}
  \label{eqPhiAlpha}
  \phi_\alpha( \mathsf{ R } ) =  \mathsf{ R }_\alpha,
\end{equation}
with $ \alpha $ a scalar parameter and $ \mathsf{ R }_\alpha $ an element of $ \mathcal{ S } $, which depends smoothly on $ \alpha $. In the above, $ \phi_\alpha $ describes a curve on the manifold (called an integral curve of $ v $), which is tangent to $ v $ at every point. 

Likewise, the pushforward $ \Phi^*( v ) $ of $ v $ is associated with a continuous transformation $ \tilde\phi_\alpha $ of intensity snapshots in manifest space. Denoting the snapshot associated with orientation $ \mathsf{ R } $ by $  a_\mathsf{ R } = \Phi( \mathsf{ R } ) $, that transformation is given by
\begin{equation}
  \label{eqPhiAlphaManifest}
  \tilde \phi_\alpha( a_\mathsf{ R } ) = a_{\mathsf{ R }_\alpha }
\end{equation} 
with $ a_{ { R }_\alpha } $ determined from Eq.~\eqref{eqPhiAlpha}. The outcome of acting on $ v $ with $ \Phi^* $ is then the directional derivative of intensity snapshots along the path defined by $ \tilde\phi_\alpha $, viz.
\begin{equation}
  \label{eqPhiV}
  \Phi^*( v ) = \lim_{\alpha\to 0} \frac{ a_{\mathsf{ R }_\alpha } - a_\mathsf{ R } }{ \alpha }.
\end{equation}

The induced metric $ g $ resulting from the above procedure will depend on the explicit form of the embedding map $ \Phi $. Here, we consider the case where $ \Phi $ describes far-field kinematic elastic scattering from a single object, where intensity amplitudes are given by the Fourier integral in Eq.~\eqref{eqDiffractionIntensity}. Other scattering scenarios can be treated in a similar manner, provided that $ \Phi $ meets the conditions of an embedding. For our purposes, it is sufficient to consider one-parameter transformations arising from left multiplication by SO(3) matrices carrying out rotations about one of the $ x $, $ y $, or $ z $ axes [see Eq.~\eqref{eqLeftRightMultiplication}]. That is, we set
\begin{equation}
  \label{eqPhiMu}
  L^\mu_\alpha( \mathsf{ R } ) = \exp( \alpha \mathsf{ J }_\mu ) \mathsf{ R }  \quad \text{for} \quad \mu \in \{ 1, 2, 3 \},
\end{equation}
where $ \alpha $ is the rotation angle in radians, and $ \mathsf{ J }_1 $, $ \mathsf{ J }_2 $, and $ \mathsf{ J }_3 $  are $ 3 \times 3 $ antisymmetric matrices generating rotations about the $ x $, $ y $, and $ z $ axes:  
\begin{equation}
  \label{eqJMu}
  \begin{gathered}
    \mathsf{ J }_1 = 
    \begin{pmatrix}
      0 & 0 & 0 \\
      0 & 0 & -1 \\
      0 & 1 & 0
    \end{pmatrix}, \quad
    \mathsf{ J }_2 = 
    \begin{pmatrix}
      0  & 0 & 1 \\
      0  & 0 & 0 \\
      -1 & 0 & 0
    \end{pmatrix}, \\
    \mathsf{ J }_3 = 
    \begin{pmatrix}
      0  & -1  & 0 \\
      1  &  0  & 0 \\
      0  &  0  & 0
    \end{pmatrix}.
  \end{gathered}
\end{equation}
We denote the vector fields generating $ \phi^\mu_\alpha $ by $ e_\mu $. It then follows from Eq.~\eqref{eqPhiV} with $ \mathsf{ R }_\alpha = L^\mu_{\alpha}( \mathsf{ R} ) $ that the pushforward fields $ \Phi^*( e_\mu ) $ are given by
\begin{equation}
  \label{eqTildeRLab}
  \Phi^*( e_\mu )( \vec{ r } ) = \left[ \omega( \vec{ r } ) \right]^{1/2} \bm{ \nabla } \bm{ \cdot } \left[ \mathsf{ J }_\mu \bm{ q } a_\mathsf{ R }( \bm{ q } ) \right] |_{\bm{q} = \bm{ q }( \vec{ r } )}.
\end{equation}  
Note that in deriving Eq.~\eqref{eqTildeRLab} we have used the divergence-free property $ \bm{ \nabla } \bm{ \cdot } \left( \mathsf{ J }_\mu \bm{ q } \right) = 0$, which applies for any scattering wavevector $ \bm{ q } $ and rotation generator $ \mathsf{ J }_\mu $.

As one may verify, the tangent vector fields $ e_\mu $ are linearly independent. This means that the set $ \{ e_1, e_2, e_3 \} $ forms a basis to expand tangent vectors on $ \mathcal{ S } $, and, in turn, the induced metric can be represented by a $ 3 \times 3 $ matrix with elements
\begin{equation}
  \label{eqGComponents}
  g_{\mu\nu}( \mathsf{ R } ) = ( \Phi^* e_\mu, \Phi^* e_\nu ).
\end{equation}
The $ g_{\mu\nu}( \mathsf{ R } ) $ above are the components of the induced metric in Eq.~\eqref{eqGExpansionR} at orientation $ \mathsf{ R } $, provided that $ E^\mu $ are dual basis vectors to $ e_\nu $, defined through the relation
\begin{equation}
  \label{eqEDual}
  E^\mu( e_\nu ) = { \delta^\mu }_\nu.
\end{equation}
Moreover, it is possible to show that $ e_\mu $ are invariant under the right multiplication map $ R_\mathsf{ Q } $ in Eq.~\eqref{eqLeftRightMultiplication} \citep[][]{Wald84}. Substituting for $ \Phi^* ( e_\mu ) $ in Eq.~\eqref{eqGComponents} using Eq.~\eqref{eqTildeRLab} then leads to the expression in Eq.~\eqref{eqGComponentsR} for the components of the induced metric in a right-invariant tensorial basis. 

Besides the form in Eq.~\eqref{eqGComponentsR}, it is useful to express the components of $ g $ in terms of spherical harmonics in reciprocal space. As is customary in diffraction theory \cite{SaldinEtAl09}, we write down the reference scattering amplitude $ a = \Phi( \mathsf{ I } ) $ corresponding to the identity matrix $ \mathsf{ I } $ using the expansion
\begin{equation}
  a( \bm{ q } ) = \sum_{j=0}^\infty \sum_{m=-j}^j a_j^m( q ) Y_j^m( \theta, \phi ), 
\end{equation}
where $ a_j^m $ are complex functions of the radial coordinate $ q $ in reciprocal space, and $ Y_j^m $ are spherical harmonics. Note the property $ a_j^{m*} = ( -1 )^m a_j^{-m} $, which is a consequence of $ a( \bm{ q } ) $ being real. Making use of the standard formula describing rotations of spherical harmonics via Wigner $ D $-matrices, 
\begin{equation}
  Y_j^m\bm{ ( } \mathsf{ R }^{-1}( \theta, \phi ) \bm{ ) } = \sum_{m'=-j}^j D^j_{mm'}( \mathsf{ R } ) Y_j^{m'}(  \theta, \phi ),
\end{equation}
it follows that the amplitude distribution $ a_\mathsf{ R } = \Phi( \mathsf{ R } ) $ corresponding to object orientation $ \mathsf{ R } $ is given by
\begin{equation}
  \label{eqA3R}
  a_\mathsf{ R }( \bm{ q } ) = \sum_{j=0}^\infty \sum_{m,m'=-j}^j a_j^{m'}( q ) D^j_{mm'}( \mathsf{ R } ) Y_j^m( \theta, \phi ).
\end{equation} 
Inserting the above in Eq.~\eqref{eqGComponentsR} leads to the following general expression for the metric components, 
\begin{multline}
  \label{eqGRAlt}
  g_{\mu\nu}( \mathsf{ R } ) = \sum_{j_1,j_2=0}^\infty \sum_{m_1,m'_1=-j_1}^{j_1} \sum_{m_2,m'_2=-j_2}^{j_2} (-1)^{m_1+m'_1} \\
  \times D^{j_1}_{-m_1,-m'_1}( \mathsf{ R } ) D^{j_2}_{m_2,m'_2}( \mathsf{ R } ) K_{\mu\nu}^{j_1m_1m'_1j_2m_2m'_2},
\end{multline}
In the above, $  K_{\mu\nu}^{j_1m_1m'_1j_2m_2m'_2} $ are complex-valued coefficients, which vanish unless the following conditions are met (together with the corresponding conditions obtained by interchanging $ \mu $ and $ \nu $):
\begin{equation}
  \label{eqYConditions}
  \begin{gathered}
    \text{$m_2 = m_1 \pm 2$ or $ m_2 = m_1 $}  \text{ for } 
    (\mu,\nu) = \begin{cases}
      (1,1),  \\
      (2,2),
    \end{cases} \\
    m_2 = m_1 \pm 2  \text{ for $ (\mu,\nu) = ( 1, 2 ) $ }, \\
    m_2 = m_1 \pm 1 \text{ for } 
    (\mu,\nu) = \begin{cases}
      (1,3), \\ 
      (2,3), 
    \end{cases}\\
    m_2 = m_1 \text{ for $ (\mu,\nu) = (3,3) $}.
  \end{gathered}
\end{equation}
    
\subsection{Decomposition into homogeneous and inhomogeneous parts}
\label{appDecomposition}

Even though the induced metric tensor $ g $ in Eq.~\eqref{eqGInduced} is not homogeneous, it is nevertheless possible to decompose it as a sum 
\begin{equation}
  g = h + w,
\end{equation}
where $ h $ is a homogeneous metric with respect to a group acting transitively on the data manifold, and $ w $ a tensor that averages to zero over the manifold. Using Fourier theory on SO(3) \cite{ChirikjianKyatkin00,KostelecRockmore03}, here we compute the components of $ h $ and $ w $ in the right-invariant basis $ E^\mu $ from Eq.~\eqref{eqEDual}. The analysis presented below yields the important results that (i) the homogeneous part $ h $ of the metric belongs to the family of metrics associated with axisymmetric rotors; (ii) the Fourier spectrum of $ w $ has a highly sparse structure, with only a limited number of nonzero coefficients contributing to metric inhomogeneity. If, as a result of the structure of its Fourier spectrum, the norm $ \lVert w \rVert $ of $ w $ (suitably defined) is smaller than the norm $ \lVert h \rVert $ of $ h $, then using the numerically-accessible eigenfunctions of $ g $ to approximate the eigenfunctions of $ h $ results in a bounded loss of accuracy. 

We begin by noting that the Fourier transform of the metric components $ g_{\mu\nu} $ in Eq.~\eqref{eqGRAlt} consists for each $ ( \mu, \nu ) $ of the sequence $ \{ \hat g^j_{\mu\nu} \}_{j=0}^\infty $ of $ ( 2 j + 1 ) \times ( 2 j + 1 ) $ matrices $ g^j_{\mu\nu} $, whose components are given by
\begin{equation}
  \label{eqFourierTransformG}
  [ \hat g^j_{\mu\nu} ]_{mm'} = \int dV ( \mathsf{ R } ) \, g_{\mu\nu}( \mathsf{ R } ) D^j_{m'm}( \mathsf{ R }^{-1} ).
\end{equation}
In the above, integration is performed over SO(3) using the volume element $ dV $ of the bi-invariant metric $ B $ in Eq.~\eqref{eqRoundMetric} \footnote{An explicit formula for the volume element associated with $ B $ (the Haar measure), expressed in terms of the $ zyz $ Euler angles $ ( \alpha^1, \alpha^2, \alpha^3 ) $ parameterizing a rotation matrix $ \mathsf{ R } $, is $ dV( \mathsf{ R } ) =  \lvert B \rvert^{1/2} d\alpha^1 \, d\alpha^2 \, d\alpha^3 $ with $ \lvert B \rvert = ( \det[ B_(\mu\nu)] )^{1/2} = \sin(\alpha^2) $.}, and $ D^j_{mm'} $ are the Wigner $ D $-functions from Eq.~\eqref{eqEigenvalueProblemD}. The collection of the $ \hat g^j_{\mu\nu} $ matrices may be used to recover $ g'( \mathsf{ R} ) $ via the inverse transform
\begin{equation}
  \label{eqInverseFourierTransformG}
  g_{\mu\nu}( \mathsf{ R } ) = \frac{ 1 }{ 8 \pi^2 } \sum_{j=0}^\infty ( 2 j + 1 ) \tr( \hat g^j_{\mu\nu} D^j( \mathsf{ R}  ) ),
\end{equation}
where $ D^j( \mathsf{ R } ) = [ D^h_{mm'}( \mathsf{ R } ) ] $ is the $ j $-th Wigner matrix of size $ ( 2 j + 1 ) \times ( 2 j + 1 ) $. Because $ D^0 $ is the trivial representation of SO(3), a constant scalar on the manifold, the term $ \hat g^0_{\mu\nu} $ corresponds to the homogeneous part of the metric. The $ \hat g^j_{\mu\nu} $ matrices with $ j \geq 1 $ give rise to the inhomogeneous tensor $ w $, which encodes object-specific information. That is, we have
\begin{equation}
  h_{\mu\nu} = \hat g^0_{\mu\nu}, \quad w_{\mu\nu} =  \frac{ 1 }{ 8 \pi^2 } \sum_{j=1}^\infty ( 2 j + 1 ) \tr( \hat g^j_{\mu\nu} D^j( \mathsf{ R}  ) ).
\end{equation}
where $ h_{\mu\nu} $ and $ w_{\mu\nu} $ are the components of $ h $ and $ w $ in the $ E^\mu $ basis, respectively.

Since each component of the metric contains a product of two Wigner $ D $-functions [see Eq.~\eqref{eqGRAlt}], the calculation of the Fourier transform is facilitated significantly by the triple-integral formula involving the 3-$jm$ symbols of quantum mechanical angular momentum \cite{VarshalovichEtAl88,ChirikjianKyatkin00}, 
\begin{multline}
  \int dV ( \mathsf{ R } ) \, D^{j*}_{mm'}( \mathsf{ R } ) D^{j_1}_{m_1m'_1}( \mathsf{ R } ) D^{j_2}_{m_2m'_2}( \mathsf{ R } ) \\
  = (-1)^{m+m'} 8 \pi^2 
  \begin{pmatrix}
    j_1 & j_2 & j \\
    m_1 & m_2 & -m 
  \end{pmatrix}
  \begin{pmatrix}
    j_1 & j_2 & j \\
    m_1' & m_2' & -m'
  \end{pmatrix}.
\end{multline}
In particular, inserting the metric components from Eq.~\eqref{eqGRAlt} in Eq.~\eqref{eqFourierTransformG}, and making use of the triple-integral formula, leads to the result
\begin{multline}
  \label{eqGHat}
  [ \hat g^j_{\mu\nu} ]_{mm'} =
  \sum_{j_1=0}^\infty \sum_{j_2=|j_1-j|}^{j_1+j} \sum_{m_1,m'_1=-j_1}^{j_1} \sum_{m_2,m'_2=-j_2}^{j_2} \\
  K_{\mu\nu}^{j_1m_1m'_1j_2m_2m'_2} \frac{ ( -1 )^{m_1+m'_1+m+m'} }{ 16 \pi^2 } \\ 
  \times
  \begin{pmatrix}
    j_1 & j_2 & j \\
    -m_1 & m_2 & m
  \end{pmatrix}
  \begin{pmatrix}
    j_1 & j_2 & j \\
    -m'_1 & m'_2 & m'
  \end{pmatrix},
\end{multline} 
where we have employed the triangle inequality, $ | j_1 -j_2 | \leq j \leq j_1 + j_2 $, obeyed by the nonzero 3-$jm $ symbols. The key point about Eq.~\eqref{eqGHat} is that the selection rules for the 3-$jm$ symbols strongly restrict the values of the azimuthal quantum number $ m $ for which $ [ \hat g^j_{\mu\nu} ]_{mm'} $ is non-vanishing. Specifically, it is possible to show that the matrix elements $ [ \hat g^j_{\mu\nu} ]_{mm'} $ are zero unless
\begin{equation}
  \begin{gathered}
    m \in \{ 0, \pm 2 \}, \quad \text{for $ ( \mu,\nu) = ( 1, 1 ) $ or $ ( \mu,\nu) = ( 2, 2) $}, \\
    m \in \{ \pm 2 \}, \quad \text{for $ ( \mu, \nu ) = ( 1, 2 ) $}, \\
    m \in \{ \pm 1 \}, \quad \text{for $ (\mu,\nu) = ( 1, 3 ) $ or $ ( \mu, \nu ) = ( 2, 3 ) $}, \\
    m \in \{ 0 \}, \quad \text{for $ (\mu,\nu) = ( 3, 3 ) $}.
  \end{gathered}
\end{equation} 
It follows that the $ j = 0 $ term of the Fourier spectrum (defined only for $ m = m' = 0 $) giving rise to the homogeneous part of the metric has no off-diagonal components. This in turn means that $ h $ can be expressed in terms of some non-negative parameters $ \ell_\mu $ in the form of Eq.~\eqref{eqHuMetric}. An explicit evaluation of the on-diagonal terms yields the additional result that the $ \ell_1 $ and $ \ell_2 $ parameters are, in fact, equal. Thus, the components of the induced metric read
\begin{equation}
  \label{eqTildeGDecomp}
  g_{\mu\nu}  = h_{\mu\nu}+ w_{\mu\nu}, \quad   
  [ h_{\mu\nu} ] = 
  \begin{pmatrix}
    \ell_{1} & 0 & 0 \\
    0 & \ell_{1} & 0 \\
    0 & 0 & \ell_{3}
  \end{pmatrix}.
\end{equation}
By construction, the average of $ w_{\mu\nu} $ over the manifold vanishes, i.e.,  
\begin{equation}
  \int dV ( \mathsf{ R } ) \, w_{\mu\nu}( \mathsf{ R } ) = 0.
\end{equation}

\subsection{Bounding the error in the eigenfunctions}
\label{appErrorBounds}

Given the above decomposition of the induced metric $ g $ into a homogeneous metric $ h $ associated with an axisymmetric rotor plus an inhomogeneous component $ w $, it is possible to write down sufficient conditions for the discrepancy between the Laplace-Beltrami eigenfunctions associated with $ g $ and $ h $ to be bounded. First, note that $ h $ induces a norm $ \lVert \cdot \rVert_{h,\infty} $ on tensors of type $ (0, 2 ) $ on the latent manifold. For a sufficiently smooth tensor field $ u $ of type $ ( 0, 2 ) $ that norm is defined as
\begin{equation}
 \lVert u \rVert_{h,\infty} = \sup_{\mathsf{R}, v} \left( \frac{ | u( v, v ) | }{ h( v, v ) } \right)^{1/2},
\end{equation}
where the supremum is taken over SO(3) matrices $ \mathsf{ R } $ and nonzero tangent vectors $ v $ on SO(3). Thus, $ \lVert u \rVert_{h,\infty} $ involves computing the least upper bound over the manifold and over the nonzero tangent vectors of the relative magnitudes of the quadratic forms $ u( v, v ) $ and $ h( v, v ) $. 

Next, set $ u = g = h + w $, and compute 
\begin{equation}
  \lVert g \rVert_{h,\infty} = \sup_{\mathsf{R}, v} \left| 1 + \frac{ w( v, v ) }{ h( v, v ) } \right|^{1/2}.
\end{equation}
By the direct and reverse triangle inequalities of norms, we have the bounds
\begin{equation}
  \label{eqWBound1}
  \left | 1 - \lVert w \rVert_{h,\infty} \right | \leq \lVert g \rVert_{h,\infty} \leq 1 + \lVert w \rVert_{h,\infty}.
\end{equation}
If $ \lVert w \rVert_{h,\infty} $ is strictly less than one, then Eq.~\eqref{eqWBound1} implies that $ \lVert g \rVert_{h,\infty} $ is bounded from below by the nonzero constant $ \Lambda_- = 1 - \lVert w \rVert_{h,\infty} $ and from above by $ \Lambda_+ = 1 + \lVert w \rVert_{h,\infty} $. 

It follows that if the condition $ \lVert w \rVert_{h,\infty} < 1 $ holds, then $ g $ and $ h $ obey the bound
\begin{equation}
  \label{eqLipshitz2}
  \frac{ 1 }{ \Lambda } \leq \lVert g \rVert_{h,\infty} \leq \Lambda 
\end{equation}
with $ \Lambda = \max\{ \Lambda_-^{-1}, \Lambda_+ \} $. This property-is called bi-Lipschitz equivalence. B\'erard \emph{et al.}\ \cite{BerardEtAl94} show that if the Ricci curvatures of two bi-Lipschitz equivalent metrics, $ g $ and $ h $, are bounded from below by some negative constant, then there exist constants $ \eta_i( \Lambda ) $, which go to zero as $ \Lambda $ approaches 1, such that for any orthonormal basis of eigenfunctions $ \{ y_i \}_{i=1}^K $ of $ \upDelta_h $, one can associate an orthonormal basis of eigenfunctions $ \{ \psi_i \}_{i=1}^K $ of $ \upDelta_g $ satisfying the bound 
\begin{equation}
  \label{eqSupNormEigenfunctionBound}
  \lVert \psi_i - y_i \rVert_\infty \leq \eta_i( \Lambda ),
\end{equation}
where $ \lVert \cdot \rVert_\infty $ denotes the supremum norm.

This means that if one approximates the set of (possibly degenerate) eigenfunctions of $ \upDelta_h $ with corresponding eigenvalues $ \lambda^h_1, \ldots, \lambda^h_K $ by the eigenfunctions of $ \upDelta_g $ with corresponding eigenvalues $ \lambda_1, \ldots, \lambda_K $ (sorted in increasing order), then the maximum error is bounded from above in a pointwise sense. Here, we do not attempt to derive conditions on the properties of the object needed to guarantee that $ \lVert w \rVert_{h,\infty} < 1 $ holds, nor do we address whether the bi-Lipschitz equivalence between $ g $ and $ h $ can be deduced by imposing weaker constraints on $ w $. Nevertheless, to the extent that the norm of $ w $ is indeed small (as suggested by, e.g., the strong selection rules on its Fourier spectrum), we expect that using the leading eigenfunctions of $ g $ for orientation recovery should produce only small systematic error, as observed in practice.

\section{Algorithms}
\label{appAlgorithms}

As an instructive application of the symmetries identified in Sec.~\ref{secTheory}, we describe an accurate and efficient algorithm for the analysis of scattering datasets. This makes explicit use of the properties of the homogeneous metric $ h $ in Eq.~\eqref{eqHuMetric} to interpret results produced by the Diffusion Map algorithm of Coifman and Lafon \cite{CoifmanLafon06}. As mentioned earlier, and also demonstrated in Paper~II, a variety of other manifold algorithms can also be used, all taking advantage of the symmetries underlying datasets produced by scattering. Here, we consider the idealized scenario of noise-free data. This will be relaxed in Paper~II, where we extend our approach to deal with snapshots severely affected by Poisson and Gaussian noise. 

Instead of a continuous data manifold $ \mathcal{ M } $, experimental data represent a countable subset $ M $ of $ \mathcal{ M } $ consisting of $ s $ identically and independently distributed (IID) samples in $ n $-dimensional data space (with $ n $ the number of detector pixels) drawn randomly from a possibly non-uniform distribution on $ \mathcal{ M } $. That is, we have 
\begin{equation}
  \label{eqDatasetA}
  M = \{ \underline{ a }_1, \ldots, \underline{ a }_s \},
\end{equation}
where $ \underline{ a }_i = ( a_{i1}, \ldots, a_{in} ) $ are $ n $-dimensional vectors of pixel amplitudes (see Fig.~\ref{figIntensity}). As described in Sec.~\ref{secRiemannianFormulation}, the amplitudes are given by $ a_{ij} = a_{\mathsf{R}_i} ( \vec{ r }_j ) $, where $ \vec{ r }_j $ is the position of pixel $ j $ in the detector plane, and $ a_{\mathsf{R}_i} = \Phi( \mathsf{ R }_i ) $ is the snapshot associated with orientation $ \mathsf{ R }_i $. 

In Ref.~\cite{CoifmanLafon06} it is shown that it is possible to construct a one-parameter family of diffusion processes (random walks) on the point cloud from Eq.~\eqref{eqDatasetA}, with each process described by an $ s \times s $ transition probability matrix $ \mathsf{ P }_\epsilon $, such that in the limit $ \epsilon \to 0 $ and $ s \to \infty $ the eigenvectors of $ \mathsf{ P }_\epsilon $ converge to the eigenfunctions of the Laplace-Beltrami operator $ \upDelta_g $ associated with the distance metric in data space [i.e., the induced metric tensor in Eq.~\eqref{eqGInduced}]. More specifically, if $ \underline{ \psi }_k $ are $ s $-dimensional column vectors in the eigenvalue problem 
\begin{equation}
  \label{eqEigenvectorsP}
  \mathsf{ P }_\epsilon \underline{ \psi }_k = \lambda_k \underline{ \psi }_k, \quad \underline{ \psi }_k = ( \psi_{1k}, \ldots, \psi_{sk} )^\text{T},
\end{equation}
then the relation $ \psi_{ik} \approx \psi_k( \mathsf{ R }_i ) $ holds for large-enough $ s $ and small-enough $ \epsilon $, where $ \psi_k( \mathsf{ R }_i ) $ is the $ k $-th eigenfunction of $ \upDelta_g $ in Eq.~\eqref{eqLapl} evaluated at element $ \mathsf{ R }_i $ on the latent manifold. 

Central to the efficiency of Diffusion Map is the fact that the transition probability matrix $ \mathsf{ P }_\epsilon $ is highly sparse. This is because $ \mathsf{ P }_\epsilon $ is constructed by suitable normalizations of a matrix $ \mathsf{ W } $ assigning weights $ W_{ij} = \mathcal{ K }( \underline{ a }_i, \underline{ a }_j ) $ via a Gaussian kernel 
\begin{equation}
  \label{eqGaussianKernel}
  \mathcal{ K }( \underline{ a }_i, \underline{ a }_j ) = \exp( - \lVert \underline{ a }_i - \underline{ a }_j \rVert^2 / \epsilon ) 
\end{equation}
to pairs of snapshots in $ M $, depending on their Euclidean distance in data space. In applications, one typically fixes a positive integer $ d $ and for each snapshot $ \underline{ a }_i $ retains the distances up to its $ d $-th nearest neighbor. Hereafter we denote the index of the $ j $-th nearest neighbor of $ \underline{ a }_i $ by $ N_{ij} $, and the corresponding distance by $ S_{ij} = \lVert \underline{ a }_i - \underline{ a }_{ N_{ij} } \rVert $. Pseudocode for computing $ \mathsf{ P }_\epsilon $ given the $ s \times d $ distance and index matrices, $ \mathsf{ S } = [ S_{ij}] $ and $ \mathsf{ N } = [ N_{ij} ] $, is listed in Table~\ref{algP}.

\begingroup 
\squeezetable
\begin{table}
  \caption{\label{algP}Calculation of the sparse transition probability matrix $ \mathsf{ P }_\epsilon $ in Diffusion Map, following Coifman and Lafon \cite{CoifmanLafon06}.}
  \begin{ruledtabular}
    \begin{tabular}{c}
      \begin{minipage}{\linewidth}
        \begin{algorithmic}[1]
          \Statex \textbf{Inputs:}
          \Statexi $ s \times d $ distance matrix $ \mathsf{ S } $
          \Statexi $ s \times d $ nearest-neighbor index matrix $ \mathsf{ N } $
          \Statexi Gaussian width $ \epsilon $
          \Statexi Normalization parameter $ \alpha $
          \Statex
          \Statex \textbf{Outputs:}
          \Statexi $ s \times s $ sparse transition probability matrix $ \mathsf{ P } $
          \Statex
          \State Construct an $ s \times s $ sparse symmetric weight matrix $ \mathsf{ W } $, such that
          \begin{displaymath}
            W_{ij} = \begin{cases}
              1, & \text{if $ i = j $}, \\
              \exp( - S_{ik}^2 / \epsilon ), &  \text{if $ j = N_{ik} $}, \\
              W_{ji}, & \text{if $ W_{ij} \neq 0$}, \\
              0, & \text{otherwise}.
            \end{cases}
          \end{displaymath}
          \State Evaluate the $ s \times s $ diagonal matrix $ \mathsf{ Q } $ with nonzero elements $ Q_{ii} = \sum_{j=1}^s W_{ij} $.
          \State Form the anisotropic kernel matrix $ \mathsf{ K } = \mathsf{ Q }^{-\alpha} \mathsf{ W } \mathsf{ Q }^{-\alpha} $.
          \State Evaluate the $ s \times s $ diagonal matrix $ \mathsf{ D } $ with nonzero elements $ D_{ii} = \sum_{j=1}^s K_{ij} $.
          \State \Return $ \mathsf{ P }_\epsilon = \mathsf{ D }^{-1} \mathsf{ K } $ 
        \end{algorithmic}
      \end{minipage}
    \end{tabular}
  \end{ruledtabular}
\end{table}  
\endgroup

Once the eigenvectors in Eq.~\eqref{eqEigenvectorsP} have been evaluated, the next step in the orientation-recovery process is to evaluate the linear-combination coefficients needed to convert the nine-dimensional embedding coordinates $ ( \psi_{l1}l, \ldots, \psi_{l9} ) $ in Eq.~\eqref{eqPsi} to elements of an approximate rotation matrix $ \tilde{ \mathsf{ R } }_l $. This conversion is performed by the discrete analog of Eq.~\eqref{eqRDCorrespondence}, namely
\begin{equation}
  \label{eqRExpansionDiscrete}
  [ \tilde{ \mathsf{ R } }_l]_{ij} = \sum_{k=1}^9 c_{ijk} \psi_{lk}l, \quad 1 \leq l \leq s.
\end{equation} 
Here the expansion coefficients $ c_{ijk} $ are determined by a least-squares minimization of the error functional
\begin{subequations}
  \label{eqNLSQ1}
  \begin{gather}
    \label{eqErrorFunctional}
    \mathcal{ G }( \{ c_{ijk} \} ) = \sum_{l=1}^{ r } G_l^2( \{ c_{ijk} \} ), \\
    G_l^2( \{ c_{ijk} \} ) = \lVert \tilde{ \mathsf{ R } }_l^\mathrm{T} \tilde{ \mathsf{ R } }_l - \mathsf{ I } \rVert^2 + | \det( \tilde{ \mathsf{ R } }_l ) - 1 |^2, \\
    [ \tilde{ \mathsf{ R } }_l ]_{ij} = \sum_{k=1}^9 c_{ijk} \psi_{lk},
  \end{gather}
\end{subequations}
where $ \lVert \mathsf{ M } \rVert = \left( \sum_{i,j=1}^3 M_{ij}^2 \right)^{1/2} $ denotes the Frobenius norm (the entrywise two-norm) of a $ 3 \times 3 $ matrix $ \mathsf{ M } = [ M_{ij} ] $. 

Because the eigenfunctions $ \psi_k $ are not exact linear combinations of the Wigner $ D $-functions in Eq.~\eqref{eqD1}, a (real) polar decomposition step may be applied to produce an exact rotation matrix $ \mathsf{ R } $, given an approximate rotation matrix estimated from the eigenfunction data via Eq.~\eqref{eqRExpansionDiscrete}. The real polar decomposition is a factorization of a real square matrix $ \tilde{ \mathsf{ R } } $ into the product
\begin{equation}
  \label{eqPolarDecomposition}
  \tilde{ \mathsf{ R } } = \mathsf{ R } \mathsf{ S }, 
\end{equation}
where $ \mathsf{ R } $ is an orthogonal matrix, and $ \mathsf{ S } $ is symmetric positive-semidefinite matrix.

The nonlinear least-squares step may be adequately performed using only a small subset of the full dataset, consisting of $ r $ datapoints drawn randomly from $ M $. For instance, in the results of Sec.~\ref{secResults}, where the total number of samples is $ s = 2 \times 10^6 $, we use $ r = 8 \times 10^4 $ datapoints. We denote the total squared residual of the optimization process by $ \mathcal{ G }^* $, where $ \mathcal{ G }^* $ is equal to $ \mathcal{ G }( \{ c_{ijk} \} ) $ in Eq.~\eqref{eqErrorFunctional}, with $ \{ c_{ijk} \} $ given by the minimizer of the error functional.

Besides $ \mathsf{ R } $, one might additionally require an evaluation of the corresponding coordinates in an SO(3) coordinate chart (e.g., for diffraction-pattern gridding). Depending on the coordinate chart of interest, various techniques are available to perform this procedure stably and efficiently. For instance, the coordinates of $ \mathsf{ R } $ in the unit-quaternion parameterization \cite{Kuipers02} can be determined by making use of the trace formula
\begin{equation}
  \label{eqTraceFormula}
  \tr( \mathsf{ R } ) = 1 + \cos\chi
\end{equation}
and the eigenvector relation
\begin{equation}
  \label{eqEigenvectorFormula}
  \mathsf{ R } \boldsymbol{ n } = \boldsymbol{ n },
\end{equation}
where $ \chi $ is the angle of the rotation represented by $ \mathsf{ R } $, and $ \boldsymbol{ n } = n_x \bm{ x } + n_y \bm{ y } + n_z \bm{ z } $ is a unit vector in $ \mathbb{ R }^3 $ directed along the corresponding axis of rotation. The unit quaternion $ \tau $ corresponding to $ \mathsf{ R }$ is given by
\begin{equation}
  \label{eqUnitQuaternion}
  \tau = ( \cos( \chi / 2 ), \sin( \chi / 2 ) n_x, \sin( \chi / 2 ) n_y, \sin( \chi / 2 ) n_z ).
\end{equation}

In outline, our orientation-recovery method consists of the following three basic steps:
\begin{enumerate}
\item Use Diffusion Map to compute the first nine non-trivial eigenfunctions $ \underline{ \psi }_1, \ldots, \underline{ \psi }_9$ of the diffusion operator $ \mathsf{ P }_\epsilon $;
  \item Determine the linear combination coefficients $ c_{ijk} $ to transform the $ \underline{ \psi }_k $ eigenvectors to (approximate) rotation matrices by solving the nonlinear least squares problem in Eq.~\eqref{eqNLSQ1}; 
  \item Use the correspondence in Eq.~\eqref{eqRExpansionDiscrete} and polar decomposition [Eq.~\eqref{eqPolarDecomposition}] to assign a rotation matrix $ \mathsf{ R }_l $ to each observed diffraction pattern, and (optionally) extract the parameters of $ \mathsf{ R }_l $ in the SO(3) parameterization of interest.
\end{enumerate}

A high-level description of the orientation-recovery procedure for noise-free data is given in Table~\ref{algNoiseFree}.

\begingroup 
\squeezetable
\begin{table}
  \caption{\label{algNoiseFree}Orientation-recovery for noise-free snapshots using Diffusion Map}
  \begin{ruledtabular}
    \begin{tabular}{c}
      \begin{minipage}{\linewidth}
        \begin{algorithmic}[1]
          \Statex \textbf{Inputs:}
          \Statexi Noise-free snapshots $ M = \{ \underline{ a }_1, \ldots, \underline{ a }_s \} $
          \Statexi Number of retained nearest neighbors $ d $ 
          \Statexi Number of datapoints in the least-squares fit, $ r $
          \Statexi Gaussian width $ \epsilon $
          \Statex
          \Statex \textbf{Outputs:}
          \Statexi Estimated quaternions $ \mathcal{ T } = \{ \tau_1, \ldots, \tau_{s} \} $
          \Statexi Nearest-neighbor index matrix $ \mathsf{ N } $  
          \Statexi Least-squares residual $\mathcal{ G }^* $
          \Statex
          \State Compute the $ s \times d $ matrices $ \mathsf{ N } $ and $ \mathsf{ S } $ such that
          \begin{displaymath}
            \begin{aligned}
            N_{ij} &= \text{index of $j$-th nearest neighbor to snapshot $ \underline{ a }_i $},\\
            S_{ij} &= \lVert \underline{ a }_i - \underline{ a }_{N_{ij}} \rVert.
            \end{aligned}
          \end{displaymath}
          \State \Return $ \mathsf{ N } $ 
          \State Compute the sparse transition probability matrix $ \mathsf{ P } $ using the algorithm in Table~\ref{algP} with inputs $ \mathsf{ S } $, $ \mathsf{ N } $, $ \epsilon $, and $ \alpha = 1 $.
          \State Solve the sparse eigenvalue problem $ \mathsf{ P } \underline{ \psi }_k = \lambda_k \underline{ \psi }_k $ for $ 0 \leq k \leq 9 $ and $ 1 = \lambda_0 < \lambda_1 \leq \cdots \leq \lambda_9 $.
          \State Solve the nonlinear least-squares problem~\eqref{eqNLSQ1}. 
          \State \Return $ \mathcal{ G }^* $
          \For{$ i = 1, \ldots, s $}
          \State Compute an approximate SO(3) matrix $ \tilde{ \mathsf{ R } }_i $ for snapshot $ \underline{ a }_i $ via Eq.~\eqref{eqRExpansionDiscrete}.
          \State Project $ \tilde{ \mathsf{ R } }_i $ to an orthogonal matrix $ \mathsf{ R }_i $ using Eq.~\eqref{eqPolarDecomposition}
          \State Convert $ \mathsf{ R }_i $ to a unit quaternion $ \tau_i $ via Eqs.~\eqref{eqTraceFormula}--\eqref{eqUnitQuaternion}.
          \State \Return $ \tau_i $
          \EndFor
        \end{algorithmic}
      \end{minipage}
    \end{tabular}
  \end{ruledtabular}
\end{table}
\endgroup

Before closing this section, we comment briefly on methods for choosing the kernel width $ \epsilon $ and the number of retained nearest neighbors $ d $. This is a common issue in manifold-embedding algorithms, and a number of strategies for determining these parameters have been developed in the literature \cite{CoifmanEtAl08,FergusonEtAl10}. Here, we determine $ \epsilon $ and $ d $ \emph{a posteriori}, by monitoring the value of the least-squares residual $ \mathcal{ G }^* $. Specifically, in the applications presented in Sec.~\ref{secResults}, we start with a number of nearest neighbors $ d \sim 100 $ and a value of $ \epsilon $ of order the mean distance to the $( d /2 )$-th nearest neighbors of the datapoints, and then perform successive refinements of these parameters seeking to minimize $ \mathcal{ G }^* $.  
  

\begin{thebibliography}{65}%
\makeatletter
\providecommand \@ifxundefined [1]{%
 \@ifx{#1\undefined}
}%
\providecommand \@ifnum [1]{%
 \ifnum #1\expandafter \@firstoftwo
 \else \expandafter \@secondoftwo
 \fi
}%
\providecommand \@ifx [1]{%
 \ifx #1\expandafter \@firstoftwo
 \else \expandafter \@secondoftwo
 \fi
}%
\providecommand \natexlab [1]{#1}%
\providecommand \enquote  [1]{``#1''}%
\providecommand \bibnamefont  [1]{#1}%
\providecommand \bibfnamefont [1]{#1}%
\providecommand \citenamefont [1]{#1}%
\providecommand \href@noop [0]{\@secondoftwo}%
\providecommand \href [0]{\begingroup \@sanitize@url \@href}%
\providecommand \@href[1]{\@@startlink{#1}\@@href}%
\providecommand \@@href[1]{\endgroup#1\@@endlink}%
\providecommand \@sanitize@url [0]{\catcode `\\12\catcode `\$12\catcode
  `\&12\catcode `\#12\catcode `\^12\catcode `\_12\catcode `\%12\relax}%
\providecommand \@@startlink[1]{}%
\providecommand \@@endlink[0]{}%
\providecommand \url  [0]{\begingroup\@sanitize@url \@url }%
\providecommand \@url [1]{\endgroup\@href {#1}{\urlprefix }}%
\providecommand \urlprefix  [0]{URL }%
\providecommand \Eprint [0]{\href }%
\providecommand \doibase [0]{http://dx.doi.org/}%
\providecommand \selectlanguage [0]{\@gobble}%
\providecommand \bibinfo  [0]{\@secondoftwo}%
\providecommand \bibfield  [0]{\@secondoftwo}%
\providecommand \translation [1]{[#1]}%
\providecommand \BibitemOpen [0]{}%
\providecommand \bibitemStop [0]{}%
\providecommand \bibitemNoStop [0]{.\EOS\space}%
\providecommand \EOS [0]{\spacefactor3000\relax}%
\providecommand \BibitemShut  [1]{\csname bibitem#1\endcsname}%
\let\auto@bib@innerbib\@empty
\bibitem [{\citenamefont {Freiwald}\ and\ \citenamefont
  {Tsao}(2010)}]{FreiwaldTsao10}%
  \BibitemOpen
  \bibfield  {author} {\bibinfo {author} {\bibfnamefont {W.~A.}\ \bibnamefont
  {Freiwald}}\ and\ \bibinfo {author} {\bibfnamefont {D.~Y.}\ \bibnamefont
  {Tsao}},\ }\href@noop {} {\bibfield  {journal} {\bibinfo  {journal}
  {Science}\ }\textbf {\bibinfo {volume} {330}},\ \bibinfo {pages} {845}
  (\bibinfo {year} {2010})}\BibitemShut {NoStop}%
\bibitem [{\citenamefont {Ross}(1951)}]{Ross51}%
  \BibitemOpen
  \bibfield  {author} {\bibinfo {author} {\bibfnamefont {W.~D.}\ \bibnamefont
  {Ross}},\ }\href@noop {} {\emph {\bibinfo {title} {Plato's Theory of
  Ideas}}}\ (\bibinfo  {publisher} {Oxford University Press},\ \bibinfo
  {address} {Oxford},\ \bibinfo {year} {1951})\BibitemShut {NoStop}%
\bibitem [{\citenamefont {Tenenbaum}\ \emph {et~al.}(2000)\citenamefont
  {Tenenbaum}, \citenamefont {de~Silva},\ and\ \citenamefont
  {Langford}}]{TenenbaumEtAl00}%
  \BibitemOpen
  \bibfield  {author} {\bibinfo {author} {\bibfnamefont {J.~B.}\ \bibnamefont
  {Tenenbaum}}, \bibinfo {author} {\bibfnamefont {V.}~\bibnamefont {de~Silva}},
  \ and\ \bibinfo {author} {\bibfnamefont {J.~C.}\ \bibnamefont {Langford}},\
  }\href@noop {} {\bibfield  {journal} {\bibinfo  {journal} {Science}\ }\textbf
  {\bibinfo {volume} {290}},\ \bibinfo {pages} {2319} (\bibinfo {year}
  {2000})}\BibitemShut {NoStop}%
\bibitem [{\citenamefont {Roweis}\ and\ \citenamefont
  {Saul}(2000)}]{RoweisSaul00}%
  \BibitemOpen
  \bibfield  {author} {\bibinfo {author} {\bibfnamefont {S.~T.}\ \bibnamefont
  {Roweis}}\ and\ \bibinfo {author} {\bibfnamefont {S.~K.}\ \bibnamefont
  {Saul}},\ }\href@noop {} {\bibfield  {journal} {\bibinfo  {journal}
  {Science}\ }\textbf {\bibinfo {volume} {290}},\ \bibinfo {pages} {2323}
  (\bibinfo {year} {2000})}\BibitemShut {NoStop}%
\bibitem [{\citenamefont {Belkin}\ and\ \citenamefont
  {Niyogi}(2003)}]{BelkinNiyogi03}%
  \BibitemOpen
  \bibfield  {author} {\bibinfo {author} {\bibfnamefont {M.}~\bibnamefont
  {Belkin}}\ and\ \bibinfo {author} {\bibfnamefont {P.}~\bibnamefont
  {Niyogi}},\ }\href@noop {} {\bibfield  {journal} {\bibinfo  {journal} {Neural
  Comput.}\ }\textbf {\bibinfo {volume} {13}},\ \bibinfo {pages} {1373}
  (\bibinfo {year} {2003})}\BibitemShut {NoStop}%
\bibitem [{\citenamefont {Donoho}\ and\ \citenamefont
  {Grimes}(2003)}]{DonohoEtAl03}%
  \BibitemOpen
  \bibfield  {author} {\bibinfo {author} {\bibfnamefont {D.~L.}\ \bibnamefont
  {Donoho}}\ and\ \bibinfo {author} {\bibfnamefont {C.}~\bibnamefont
  {Grimes}},\ }\href@noop {} {\bibfield  {journal} {\bibinfo  {journal} {Proc.
  Natl. Acad. Sci.}\ }\textbf {\bibinfo {volume} {100}},\ \bibinfo {pages}
  {5591} (\bibinfo {year} {2003})}\BibitemShut {NoStop}%
\bibitem [{\citenamefont {Coifman}\ \emph {et~al.}(2005)\citenamefont
  {Coifman}, \citenamefont {Lafon}, \citenamefont {Lee}, \citenamefont
  {Maggioni}, \citenamefont {Nadler}, \citenamefont {Warner},\ and\
  \citenamefont {Zucker}}]{CoifmanEtAl05}%
  \BibitemOpen
  \bibfield  {author} {\bibinfo {author} {\bibfnamefont {R.~R.}\ \bibnamefont
  {Coifman}}, \bibinfo {author} {\bibfnamefont {S.}~\bibnamefont {Lafon}},
  \bibinfo {author} {\bibfnamefont {A.~B.}\ \bibnamefont {Lee}}, \bibinfo
  {author} {\bibfnamefont {M.}~\bibnamefont {Maggioni}}, \bibinfo {author}
  {\bibfnamefont {B.}~\bibnamefont {Nadler}}, \bibinfo {author} {\bibfnamefont
  {F.}~\bibnamefont {Warner}}, \ and\ \bibinfo {author} {\bibfnamefont
  {S.}~\bibnamefont {Zucker}},\ }\href@noop {} {\bibfield  {journal} {\bibinfo
  {journal} {Proc. Natl. Acad. Sci.}\ }\textbf {\bibinfo {volume} {102}},\
  \bibinfo {pages} {7426} (\bibinfo {year} {2005})}\BibitemShut {NoStop}%
\bibitem [{\citenamefont {Coifman}\ and\ \citenamefont
  {Lafon}(2006)}]{CoifmanLafon06}%
  \BibitemOpen
  \bibfield  {author} {\bibinfo {author} {\bibfnamefont {R.~R.}\ \bibnamefont
  {Coifman}}\ and\ \bibinfo {author} {\bibfnamefont {S.}~\bibnamefont
  {Lafon}},\ }\href@noop {} {\bibfield  {journal} {\bibinfo  {journal} {Appl.
  Comput. Harmon. Anal.}\ }\textbf {\bibinfo {volume} {21}},\ \bibinfo {pages}
  {5} (\bibinfo {year} {2006})}\BibitemShut {NoStop}%
\bibitem [{\citenamefont {Coifman}\ \emph {et~al.}(2010)\citenamefont
  {Coifman}, \citenamefont {Shkolnisky}, \citenamefont {Sigworth},\ and\
  \citenamefont {Singer}}]{CoifmanEtAl10}%
  \BibitemOpen
  \bibfield  {author} {\bibinfo {author} {\bibfnamefont {R.~R.}\ \bibnamefont
  {Coifman}}, \bibinfo {author} {\bibfnamefont {Y.}~\bibnamefont {Shkolnisky}},
  \bibinfo {author} {\bibfnamefont {F.~J.}\ \bibnamefont {Sigworth}}, \ and\
  \bibinfo {author} {\bibfnamefont {A.}~\bibnamefont {Singer}},\ }\href@noop {}
  {\bibfield  {journal} {\bibinfo  {journal} {Appl. Comput. Harmon. Anal.}\
  }\textbf {\bibinfo {volume} {28}},\ \bibinfo {pages} {296} (\bibinfo {year}
  {2010})}\BibitemShut {NoStop}%
\bibitem [{\citenamefont {Ferguson}\ \emph {et~al.}(2010)\citenamefont
  {Ferguson}, \citenamefont {Panagiotopoulos}, \citenamefont {Debenedetti},\
  and\ \citenamefont {Kevrekidis}}]{FergusonEtAl10}%
  \BibitemOpen
  \bibfield  {author} {\bibinfo {author} {\bibfnamefont {A.~L.}\ \bibnamefont
  {Ferguson}}, \bibinfo {author} {\bibfnamefont {A.~Z.}\ \bibnamefont
  {Panagiotopoulos}}, \bibinfo {author} {\bibfnamefont {P.~G.}\ \bibnamefont
  {Debenedetti}}, \ and\ \bibinfo {author} {\bibfnamefont {I.~G.}\ \bibnamefont
  {Kevrekidis}},\ }\href@noop {} {\bibfield  {journal} {\bibinfo  {journal}
  {Proc. Natl. Acad. Sci.}\ }\textbf {\bibinfo {volume} {107}},\ \bibinfo
  {pages} {13597} (\bibinfo {year} {2010})}\BibitemShut {NoStop}%
\bibitem [{\citenamefont {Singer}\ \emph {et~al.}(2010)\citenamefont {Singer},
  \citenamefont {Coifman}, \citenamefont {Sigworth}, \citenamefont {Chester},\
  and\ \citenamefont {Shkolnisky}}]{SingerEtAl10}%
  \BibitemOpen
  \bibfield  {author} {\bibinfo {author} {\bibfnamefont {A.}~\bibnamefont
  {Singer}}, \bibinfo {author} {\bibfnamefont {R.~R.}\ \bibnamefont {Coifman}},
  \bibinfo {author} {\bibfnamefont {F.~J.}\ \bibnamefont {Sigworth}}, \bibinfo
  {author} {\bibfnamefont {D.~W.}\ \bibnamefont {Chester}}, \ and\ \bibinfo
  {author} {\bibfnamefont {Y.}~\bibnamefont {Shkolnisky}},\ }\href@noop {}
  {\bibfield  {journal} {\bibinfo  {journal} {J. Struct. Biol.}\ }\textbf
  {\bibinfo {volume} {169}},\ \bibinfo {pages} {312} (\bibinfo {year}
  {2010})}\BibitemShut {NoStop}%
\bibitem [{EPA()}]{EPAPS}%
  \BibitemOpen
  \href@noop {} {}\bibinfo {note} {See EPAPS Document No.\ [] for a movie of 3D
  reconstruction. For more information on EPAPS, see
  \url{http://www.aip.org/pubservs/epaps.html.}}\BibitemShut {Stop}%
\bibitem [{\citenamefont {Schwander}\ \emph
  {et~al.}(2011{\natexlab{a}})\citenamefont {Schwander}, \citenamefont
  {Giannakis}, \citenamefont {Yoon},\ and\ \citenamefont
  {Ourmazd}}]{GiannakisEtAl11b}%
  \BibitemOpen
  \bibfield  {author} {\bibinfo {author} {\bibfnamefont {P.}~\bibnamefont
  {Schwander}}, \bibinfo {author} {\bibfnamefont {D.}~\bibnamefont
  {Giannakis}}, \bibinfo {author} {\bibfnamefont {C.~H.}\ \bibnamefont {Yoon}},
  \ and\ \bibinfo {author} {\bibfnamefont {A.}~\bibnamefont {Ourmazd}},\
  }\href@noop {} {\bibfield  {journal} {\bibinfo  {journal} {Phys. Rev. E}\ }
  (\bibinfo {year} {2011}{\natexlab{a}})},\ \bibinfo {note}
  {submitted}\BibitemShut {NoStop}%
\bibitem [{\citenamefont {Gerchberg}\ and\ \citenamefont
  {Saxton}(1972)}]{GerchbergSaxton72}%
  \BibitemOpen
  \bibfield  {author} {\bibinfo {author} {\bibfnamefont {R.~W.}\ \bibnamefont
  {Gerchberg}}\ and\ \bibinfo {author} {\bibfnamefont {W.~O.}\ \bibnamefont
  {Saxton}},\ }\href@noop {} {\bibfield  {journal} {\bibinfo  {journal}
  {Optik}\ }\textbf {\bibinfo {volume} {35}},\ \bibinfo {pages} {237} (\bibinfo
  {year} {1972})}\BibitemShut {NoStop}%
\bibitem [{\citenamefont {Fienup}(1978)}]{Fienup78}%
  \BibitemOpen
  \bibfield  {author} {\bibinfo {author} {\bibfnamefont {J.~R.}\ \bibnamefont
  {Fienup}},\ }\href@noop {} {\bibfield  {journal} {\bibinfo  {journal} {Opt.
  Lett.}\ }\textbf {\bibinfo {volume} {3}},\ \bibinfo {pages} {27} (\bibinfo
  {year} {1978})}\BibitemShut {NoStop}%
\bibitem [{\citenamefont {Oszl\'anyi}\ and\ \citenamefont
  {S\"{u}to}(2004)}]{OszlanyiSuto04}%
  \BibitemOpen
  \bibfield  {author} {\bibinfo {author} {\bibfnamefont {F.}~\bibnamefont
  {Oszl\'anyi}}\ and\ \bibinfo {author} {\bibfnamefont {A.}~\bibnamefont
  {S\"{u}to}},\ }\href@noop {} {\bibfield  {journal} {\bibinfo  {journal} {Acta
  Cryst. A}\ }\textbf {\bibinfo {volume} {60}},\ \bibinfo {pages} {134}
  (\bibinfo {year} {2004})}\BibitemShut {NoStop}%
\bibitem [{\citenamefont {Natterer}(2001)}]{Natterer01}%
  \BibitemOpen
  \bibfield  {author} {\bibinfo {author} {\bibfnamefont {F.}~\bibnamefont
  {Natterer}},\ }\href@noop {} {\emph {\bibinfo {title} {The Mathematics of
  Computerized Tomography}}}\ (\bibinfo  {publisher} {SIAM},\ \bibinfo
  {address} {Philadelphia},\ \bibinfo {year} {2001})\BibitemShut {NoStop}%
\bibitem [{\citenamefont {Frank}(2002)}]{Frank02}%
  \BibitemOpen
  \bibfield  {author} {\bibinfo {author} {\bibfnamefont {J.}~\bibnamefont
  {Frank}},\ }\href@noop {} {\bibfield  {journal} {\bibinfo  {journal} {Annu.
  Rev. Biophys. Biomolec. Struct.}\ }\textbf {\bibinfo {volume} {31}},\
  \bibinfo {pages} {303} (\bibinfo {year} {2002})}\BibitemShut {NoStop}%
\bibitem [{\citenamefont {Fung}\ \emph {et~al.}(2008)\citenamefont {Fung},
  \citenamefont {Shneerson}, \citenamefont {Saldin},\ and\ \citenamefont
  {Ourmazd}}]{FungEtAl08}%
  \BibitemOpen
  \bibfield  {author} {\bibinfo {author} {\bibfnamefont {R.}~\bibnamefont
  {Fung}}, \bibinfo {author} {\bibfnamefont {V.}~\bibnamefont {Shneerson}},
  \bibinfo {author} {\bibfnamefont {D.~K.}\ \bibnamefont {Saldin}}, \ and\
  \bibinfo {author} {\bibfnamefont {A.}~\bibnamefont {Ourmazd}},\ }\href
  {\doibase 10.1038/nphys1129} {\bibfield  {journal} {\bibinfo  {journal} {Nat.
  Phys.}\ }\textbf {\bibinfo {volume} {5}},\ \bibinfo {pages} {64} (\bibinfo
  {year} {2008})}\BibitemShut {NoStop}%
\bibitem [{\citenamefont {Loh}\ and\ \citenamefont {Elser}(2009)}]{LohElser09}%
  \BibitemOpen
  \bibfield  {author} {\bibinfo {author} {\bibfnamefont {N.~T.~D.}\
  \bibnamefont {Loh}}\ and\ \bibinfo {author} {\bibfnamefont {V.}~\bibnamefont
  {Elser}},\ }\href@noop {} {\bibfield  {journal} {\bibinfo  {journal} {Phys.
  Rev. E}\ }\textbf {\bibinfo {volume} {80}},\ \bibinfo {pages} {026705}
  (\bibinfo {year} {2009})}\BibitemShut {NoStop}%
\bibitem [{\citenamefont {Schwander}\ \emph {et~al.}(2010)\citenamefont
  {Schwander}, \citenamefont {Fung}, \citenamefont {Phillips},\ and\
  \citenamefont {Ourmazd}}]{SchwanderEtAl10a}%
  \BibitemOpen
  \bibfield  {author} {\bibinfo {author} {\bibfnamefont {P.}~\bibnamefont
  {Schwander}}, \bibinfo {author} {\bibfnamefont {R.}~\bibnamefont {Fung}},
  \bibinfo {author} {\bibfnamefont {G.~N.}\ \bibnamefont {Phillips}}, \ and\
  \bibinfo {author} {\bibfnamefont {A.}~\bibnamefont {Ourmazd}},\ }\href@noop
  {} {\bibfield  {journal} {\bibinfo  {journal} {New J. Phys.}\ }\textbf
  {\bibinfo {volume} {12}},\ \bibinfo {pages} {1} (\bibinfo {year}
  {2010})}\BibitemShut {NoStop}%
\bibitem [{\citenamefont {Scheres}\ \emph {et~al.}(2007)\citenamefont {Scheres}
  \emph {et~al.}}]{ScheresEtAl07}%
  \BibitemOpen
  \bibfield  {author} {\bibinfo {author} {\bibfnamefont {S.~H.~W.}\
  \bibnamefont {Scheres}} \emph {et~al.},\ }\href {\doibase 10.1038/nmeth992}
  {\bibfield  {journal} {\bibinfo  {journal} {Nature Methods}\ }\textbf
  {\bibinfo {volume} {4}},\ \bibinfo {pages} {27} (\bibinfo {year}
  {2007})}\BibitemShut {NoStop}%
\bibitem [{\citenamefont {Fischer}\ \emph {et~al.}(2010)\citenamefont
  {Fischer}, \citenamefont {Konevega}, \citenamefont {Wintermeyer},
  \citenamefont {Rodnina},\ and\ \citenamefont {Stark}}]{FischerEtAl10}%
  \BibitemOpen
  \bibfield  {author} {\bibinfo {author} {\bibfnamefont {N.}~\bibnamefont
  {Fischer}}, \bibinfo {author} {\bibfnamefont {A.~L.}\ \bibnamefont
  {Konevega}}, \bibinfo {author} {\bibfnamefont {W.}~\bibnamefont
  {Wintermeyer}}, \bibinfo {author} {\bibfnamefont {M.~V.}\ \bibnamefont
  {Rodnina}}, \ and\ \bibinfo {author} {\bibfnamefont {H.}~\bibnamefont
  {Stark}},\ }\href@noop {} {\bibfield  {journal} {\bibinfo  {journal}
  {Nature}\ }\textbf {\bibinfo {volume} {466}},\ \bibinfo {pages} {329}
  (\bibinfo {year} {2010})}\BibitemShut {NoStop}%
\bibitem [{\citenamefont {Younes}\ \emph {et~al.}(2008)\citenamefont {Younes},
  \citenamefont {Michor}, \citenamefont {Shah},\ and\ \citenamefont
  {Mumford}}]{YounesEtAl08}%
  \BibitemOpen
  \bibfield  {author} {\bibinfo {author} {\bibfnamefont {L.}~\bibnamefont
  {Younes}}, \bibinfo {author} {\bibfnamefont {P.~W.}\ \bibnamefont {Michor}},
  \bibinfo {author} {\bibfnamefont {J.}~\bibnamefont {Shah}}, \ and\ \bibinfo
  {author} {\bibfnamefont {D.}~\bibnamefont {Mumford}},\ }\href@noop {}
  {\bibfield  {journal} {\bibinfo  {journal} {Rend. Lincei Mat. Appl.}\
  }\textbf {\bibinfo {volume} {9}},\ \bibinfo {pages} {25} (\bibinfo {year}
  {2008})}\BibitemShut {NoStop}%
\bibitem [{\citenamefont {Lin}\ \emph {et~al.}(2006)\citenamefont {Lin},
  \citenamefont {Zha},\ and\ \citenamefont {Lee}}]{LinEtAl06}%
  \BibitemOpen
  \bibfield  {author} {\bibinfo {author} {\bibfnamefont {T.}~\bibnamefont
  {Lin}}, \bibinfo {author} {\bibfnamefont {H.}~\bibnamefont {Zha}}, \ and\
  \bibinfo {author} {\bibfnamefont {S.}~\bibnamefont {Lee}},\ }in\ \href@noop
  {} {\emph {\bibinfo {booktitle} {ECCV Part I, LNCS}}},\ \bibinfo {editor}
  {edited by\ \bibinfo {editor} {\bibfnamefont {A.}~\bibnamefont {Leondardis}},
  \bibinfo {editor} {\bibfnamefont {H.}~\bibnamefont {Bischof}}, \ and\
  \bibinfo {editor} {\bibfnamefont {A.}~\bibnamefont {Pinz}}}\ (\bibinfo
  {publisher} {Springer Verlag},\ \bibinfo {address} {Berlin, Heidelberg},\
  \bibinfo {year} {2006})\ pp.\ \bibinfo {pages} {44--55}\BibitemShut {NoStop}%
\bibitem [{\citenamefont {Sch\"olkopf}\ \emph {et~al.}(1998)\citenamefont
  {Sch\"olkopf}, \citenamefont {Smola},\ and\ \citenamefont
  {M\"uller}}]{SchoelkopfEtAl98}%
  \BibitemOpen
  \bibfield  {author} {\bibinfo {author} {\bibfnamefont {B.}~\bibnamefont
  {Sch\"olkopf}}, \bibinfo {author} {\bibfnamefont {B.}~\bibnamefont {Smola}},
  \ and\ \bibinfo {author} {\bibfnamefont {K.-R.}\ \bibnamefont {M\"uller}},\
  }\href@noop {} {\bibfield  {journal} {\bibinfo  {journal} {Neural
  Computation}\ }\textbf {\bibinfo {volume} {10}},\ \bibinfo {pages} {1299}
  (\bibinfo {year} {1998})}\BibitemShut {NoStop}%
\bibitem [{\citenamefont {Bishop}\ \emph {et~al.}(1998)\citenamefont {Bishop},
  \citenamefont {Svensen},\ and\ \citenamefont {Williams}}]{BishopEtAl98}%
  \BibitemOpen
  \bibfield  {author} {\bibinfo {author} {\bibfnamefont {C.~M.}\ \bibnamefont
  {Bishop}}, \bibinfo {author} {\bibfnamefont {M.}~\bibnamefont {Svensen}}, \
  and\ \bibinfo {author} {\bibfnamefont {C.~K.~I.}\ \bibnamefont {Williams}},\
  }\href@noop {} {\bibfield  {journal} {\bibinfo  {journal} {Neural
  Computation}\ }\textbf {\bibinfo {volume} {463}},\ \bibinfo {pages} {379}
  (\bibinfo {year} {1998})}\BibitemShut {NoStop}%
\bibitem [{\citenamefont {Moths}\ and\ \citenamefont
  {Ourmazd}(2011)}]{MothsOurmazd11}%
  \BibitemOpen
  \bibfield  {author} {\bibinfo {author} {\bibfnamefont {B.}~\bibnamefont
  {Moths}}\ and\ \bibinfo {author} {\bibfnamefont {A.}~\bibnamefont
  {Ourmazd}},\ }\href@noop {} {\bibfield  {journal} {\bibinfo  {journal} {Acta.
  Cryst.}\ } (\bibinfo {year} {2011})},\ \bibinfo {note} {in press}\BibitemShut
  {NoStop}%
\bibitem [{\citenamefont {Balasubramanian}\ and\ \citenamefont
  {Schwartz}(2002)}]{BalasubramanianSchwartz02}%
  \BibitemOpen
  \bibfield  {author} {\bibinfo {author} {\bibfnamefont {M.}~\bibnamefont
  {Balasubramanian}}\ and\ \bibinfo {author} {\bibfnamefont {E.~L.}\
  \bibnamefont {Schwartz}},\ }\href@noop {} {\bibfield  {journal} {\bibinfo
  {journal} {Science}\ }\textbf {\bibinfo {volume} {295}},\ \bibinfo {pages}
  {5552} (\bibinfo {year} {2002})}\BibitemShut {NoStop}%
\bibitem [{\citenamefont {Coifman}\ \emph {et~al.}(2008)\citenamefont
  {Coifman}, \citenamefont {Shkolnisky}, \citenamefont {Sigworth},\ and\
  \citenamefont {Singer}}]{CoifmanEtAl08}%
  \BibitemOpen
  \bibfield  {author} {\bibinfo {author} {\bibfnamefont {R.}~\bibnamefont
  {Coifman}}, \bibinfo {author} {\bibfnamefont {Y.}~\bibnamefont {Shkolnisky}},
  \bibinfo {author} {\bibfnamefont {F.}~\bibnamefont {Sigworth}}, \ and\
  \bibinfo {author} {\bibfnamefont {A.}~\bibnamefont {Singer}},\ }\href@noop {}
  {\bibfield  {journal} {\bibinfo  {journal} {IEEE Trans. Image Process.}\
  }\textbf {\bibinfo {volume} {17}},\ \bibinfo {pages} {1891} (\bibinfo {year}
  {2008})}\BibitemShut {NoStop}%
\bibitem [{\citenamefont {Schutz}(1980)}]{Schutz80}%
  \BibitemOpen
  \bibfield  {author} {\bibinfo {author} {\bibfnamefont {B.~F.}\ \bibnamefont
  {Schutz}},\ }\href@noop {} {\emph {\bibinfo {title} {Geometrical Methods of
  Mathematical Physics}}}\ (\bibinfo  {publisher} {Cambridge University
  Press},\ \bibinfo {address} {Cambridge},\ \bibinfo {year} {1980})\BibitemShut
  {NoStop}%
\bibitem [{\citenamefont {Lang}(2002)}]{Lang02}%
  \BibitemOpen
  \bibfield  {author} {\bibinfo {author} {\bibfnamefont {S.}~\bibnamefont
  {Lang}},\ }\href@noop {} {\emph {\bibinfo {title} {Introduction to
  Differentiable Manifolds}}}\ (\bibinfo  {publisher} {Springer-Verlag},\
  \bibinfo {address} {New York},\ \bibinfo {year} {2002})\BibitemShut {NoStop}%
\bibitem [{\citenamefont {Wigner}(1959)}]{Wigner59}%
  \BibitemOpen
  \bibfield  {author} {\bibinfo {author} {\bibfnamefont {E.~P.}\ \bibnamefont
  {Wigner}},\ }\href@noop {} {\emph {\bibinfo {title} {Group Theory and its
  Application to the Quantum Mechanics of Atomic Spectra}}}\ (\bibinfo
  {publisher} {Academic Press},\ \bibinfo {address} {New York},\ \bibinfo
  {year} {1959})\BibitemShut {NoStop}%
\bibitem [{\citenamefont {Biedenharn}\ and\ \citenamefont
  {Louck}(1981)}]{BiedenharnLouck81}%
  \BibitemOpen
  \bibfield  {author} {\bibinfo {author} {\bibfnamefont {L.~C.}\ \bibnamefont
  {Biedenharn}}\ and\ \bibinfo {author} {\bibfnamefont {J.~D.}\ \bibnamefont
  {Louck}},\ }\href@noop {} {\emph {\bibinfo {title} {Angular Momentum in
  Quantum Physics}}}\ (\bibinfo  {publisher} {Addison Wesley},\ \bibinfo
  {address} {Reading},\ \bibinfo {year} {1981})\BibitemShut {NoStop}%
\bibitem [{\citenamefont {Chirikjian}\ and\ \citenamefont
  {Kyatkin}(2000)}]{ChirikjianKyatkin00}%
  \BibitemOpen
  \bibfield  {author} {\bibinfo {author} {\bibfnamefont {G.~S.}\ \bibnamefont
  {Chirikjian}}\ and\ \bibinfo {author} {\bibfnamefont {A.~B.}\ \bibnamefont
  {Kyatkin}},\ }\href@noop {} {\emph {\bibinfo {title} {Engineering
  Applications of Noncummutative Harmonic Analysis: With Emphasis on Rotation
  and Motion Groups}}}\ (\bibinfo  {publisher} {CRC Press},\ \bibinfo {address}
  {Boca Raton},\ \bibinfo {year} {2000})\BibitemShut {NoStop}%
\bibitem [{\citenamefont {Hu}(1973)}]{Hu73}%
  \BibitemOpen
  \bibfield  {author} {\bibinfo {author} {\bibfnamefont {B.~L.}\ \bibnamefont
  {Hu}},\ }\href@noop {} {\bibfield  {journal} {\bibinfo  {journal} {Phys. Rev.
  D}\ }\textbf {\bibinfo {volume} {8}},\ \bibinfo {pages} {1048} (\bibinfo
  {year} {1973})}\BibitemShut {NoStop}%
\bibitem [{\citenamefont {Taub}(1951)}]{Taub51}%
  \BibitemOpen
  \bibfield  {author} {\bibinfo {author} {\bibfnamefont {A.~H.}\ \bibnamefont
  {Taub}},\ }\href@noop {} {\bibfield  {journal} {\bibinfo  {journal} {Ann.
  Math.}\ }\textbf {\bibinfo {volume} {53}},\ \bibinfo {pages} {472} (\bibinfo
  {year} {1951})}\BibitemShut {NoStop}%
\bibitem [{\citenamefont {Cowley}(1995)}]{Cowley95}%
  \BibitemOpen
  \bibfield  {author} {\bibinfo {author} {\bibfnamefont {J.~M.}\ \bibnamefont
  {Cowley}},\ }\href@noop {} {\emph {\bibinfo {title} {Diffraction Physics}}},\
  \bibinfo {edition} {3rd}\ ed.\ (\bibinfo  {publisher} {North Holland},\
  \bibinfo {address} {Amsterdam},\ \bibinfo {year} {1995})\ p.\ \bibinfo
  {pages} {496}\BibitemShut {NoStop}%
\bibitem [{\citenamefont {Arvanitoyeorgos}(2003)}]{Arvanitoyeorgos03}%
  \BibitemOpen
  \bibfield  {author} {\bibinfo {author} {\bibfnamefont {A.}~\bibnamefont
  {Arvanitoyeorgos}},\ }\href@noop {} {\emph {\bibinfo {title} {An Introduction
  to {L}ie Groups and the Geometry of Homogeneous Spaces}}}\ (\bibinfo
  {publisher} {American Mathematical Society},\ \bibinfo {address}
  {Providence},\ \bibinfo {year} {2003})\BibitemShut {NoStop}%
\bibitem [{\citenamefont {Kuipers}(2002)}]{Kuipers02}%
  \BibitemOpen
  \bibfield  {author} {\bibinfo {author} {\bibfnamefont {J.~B.}\ \bibnamefont
  {Kuipers}},\ }\href@noop {} {\emph {\bibinfo {title} {Quaternions and
  Rotation Sequences: A Primer with Applications to Orbits, Aerospace, and
  Virtual Reality}}}\ (\bibinfo  {publisher} {Princeton University Press},\
  \bibinfo {address} {Princeton},\ \bibinfo {year} {2002})\BibitemShut
  {NoStop}%
\bibitem [{\citenamefont {Lang}(1998)}]{Lang98}%
  \BibitemOpen
  \bibfield  {author} {\bibinfo {author} {\bibfnamefont {S.}~\bibnamefont
  {Lang}},\ }\href@noop {} {\emph {\bibinfo {title} {Fundamentals of
  Differential Geometry}}}\ (\bibinfo  {publisher} {Springer-Verlag},\ \bibinfo
  {address} {New York},\ \bibinfo {year} {1998})\BibitemShut {NoStop}%
\bibitem [{\citenamefont {Bronstein}\ \emph {et~al.}(2007)\citenamefont
  {Bronstein}, \citenamefont {Bronstein},\ and\ \citenamefont
  {Kimmel}}]{BronsteinEtAl07}%
  \BibitemOpen
  \bibfield  {author} {\bibinfo {author} {\bibfnamefont {A.~M.}\ \bibnamefont
  {Bronstein}}, \bibinfo {author} {\bibfnamefont {M.~M.}\ \bibnamefont
  {Bronstein}}, \ and\ \bibinfo {author} {\bibfnamefont {R.}~\bibnamefont
  {Kimmel}},\ }\href@noop {} {\emph {\bibinfo {title} {Numerical Geometry of
  Non-Rigid Shapes}}}\ (\bibinfo  {publisher} {Springer},\ \bibinfo {year}
  {2007})\BibitemShut {NoStop}%
\bibitem [{\citenamefont {Sauer}\ \emph {et~al.}(1991)\citenamefont {Sauer},
  \citenamefont {Yorke},\ and\ \citenamefont {Casdagli}}]{SauerEtAl91}%
  \BibitemOpen
  \bibfield  {author} {\bibinfo {author} {\bibfnamefont {T.}~\bibnamefont
  {Sauer}}, \bibinfo {author} {\bibfnamefont {J.~A.}\ \bibnamefont {Yorke}}, \
  and\ \bibinfo {author} {\bibfnamefont {M.}~\bibnamefont {Casdagli}},\
  }\href@noop {} {\bibfield  {journal} {\bibinfo  {journal} {J. Stat. Phys.}\
  }\textbf {\bibinfo {volume} {65}},\ \bibinfo {pages} {579} (\bibinfo {year}
  {1991})}\BibitemShut {NoStop}%
\bibitem [{\citenamefont {Wald}(1984)}]{Wald84}%
  \BibitemOpen
  \bibfield  {author} {\bibinfo {author} {\bibfnamefont {R.~M.}\ \bibnamefont
  {Wald}},\ }\href@noop {} {\emph {\bibinfo {title} {General Relativity}}}\
  (\bibinfo  {publisher} {The University of Chicago Press},\ \bibinfo {address}
  {Chicago},\ \bibinfo {year} {1984})\BibitemShut {NoStop}%
\bibitem [{\citenamefont {B\'erard}(1989)}]{Berard86}%
  \BibitemOpen
  \bibfield  {author} {\bibinfo {author} {\bibfnamefont {P.~H.}\ \bibnamefont
  {B\'erard}},\ }\href@noop {} {\emph {\bibinfo {title} {Spectral Geometry:
  Direct and Inverse Problems}}},\ \bibinfo {series} {Lecture Notes in
  Mathematics}, Vol.\ \bibinfo {volume} {1207}\ (\bibinfo  {publisher}
  {Springer-Verlag},\ \bibinfo {address} {Berlin},\ \bibinfo {year}
  {1989})\BibitemShut {NoStop}%
\bibitem [{\citenamefont {Vilenkin}(1968)}]{Vilenkin68}%
  \BibitemOpen
  \bibfield  {author} {\bibinfo {author} {\bibfnamefont {N.~J.}\ \bibnamefont
  {Vilenkin}},\ }\href@noop {} {\emph {\bibinfo {title} {Special Functions and
  the Theory of Group Representations}}},\ \bibinfo {series} {Mathematical
  Monographs}, Vol.~\bibinfo {volume} {22}\ (\bibinfo  {publisher} {American
  Mathematical Society},\ \bibinfo {address} {Providence},\ \bibinfo {year}
  {1968})\BibitemShut {NoStop}%
\bibitem [{\citenamefont {Varshalovich}\ \emph {et~al.}(1988)\citenamefont
  {Varshalovich}, \citenamefont {Moskalev},\ and\ \citenamefont
  {Khersonskii}}]{VarshalovichEtAl88}%
  \BibitemOpen
  \bibfield  {author} {\bibinfo {author} {\bibfnamefont {D.~A.}\ \bibnamefont
  {Varshalovich}}, \bibinfo {author} {\bibfnamefont {A.~N.}\ \bibnamefont
  {Moskalev}}, \ and\ \bibinfo {author} {\bibfnamefont {V.~K.}\ \bibnamefont
  {Khersonskii}},\ }\href@noop {} {\emph {\bibinfo {title} {Quantum Theory of
  Angular Momentum}}}\ (\bibinfo  {publisher} {World Scientific},\ \bibinfo
  {address} {Singapore},\ \bibinfo {year} {1988})\BibitemShut {NoStop}%
\bibitem [{\citenamefont {McCauley}(1997)}]{McCauley97}%
  \BibitemOpen
  \bibfield  {author} {\bibinfo {author} {\bibfnamefont {J.~L.}\ \bibnamefont
  {McCauley}},\ }\href@noop {} {\emph {\bibinfo {title} {Classical Mechanics:
  Transformations, Flows, Integrable and Chaotic Dynamics}}}\ (\bibinfo
  {publisher} {Cambridge University Press},\ \bibinfo {address} {Cambridge},\
  \bibinfo {year} {1997})\BibitemShut {NoStop}%
\bibitem [{\citenamefont {Misner}(1969)}]{Misner69}%
  \BibitemOpen
  \bibfield  {author} {\bibinfo {author} {\bibfnamefont {C.}~\bibnamefont
  {Misner}},\ }\href@noop {} {\bibfield  {journal} {\bibinfo  {journal} {Phy.
  Rev. Lett.}\ }\textbf {\bibinfo {volume} {22}},\ \bibinfo {pages} {1071}
  (\bibinfo {year} {1969})}\BibitemShut {NoStop}%
\bibitem [{\citenamefont {B\'erard}\ \emph {et~al.}(1994)\citenamefont
  {B\'erard}, \citenamefont {Besson},\ and\ \citenamefont
  {Gallot}}]{BerardEtAl94}%
  \BibitemOpen
  \bibfield  {author} {\bibinfo {author} {\bibfnamefont {P.}~\bibnamefont
  {B\'erard}}, \bibinfo {author} {\bibfnamefont {G.}~\bibnamefont {Besson}}, \
  and\ \bibinfo {author} {\bibfnamefont {S.}~\bibnamefont {Gallot}},\
  }\href@noop {} {\bibfield  {journal} {\bibinfo  {journal} {Geom. Funct.
  Anal.}\ }\textbf {\bibinfo {volume} {4}},\ \bibinfo {pages} {373} (\bibinfo
  {year} {1994})}\BibitemShut {NoStop}%
\bibitem [{\citenamefont {Cao}\ and\ \citenamefont {Zhu}(2006)}]{CaoZhu06}%
  \BibitemOpen
  \bibfield  {author} {\bibinfo {author} {\bibfnamefont {H.-D.}\ \bibnamefont
  {Cao}}\ and\ \bibinfo {author} {\bibfnamefont {X.-P.}\ \bibnamefont {Zhu}},\
  }\href@noop {} {\enquote {\bibinfo {title} {Hamilton-{P}erelman's proof of
  the {P}oincar\'e conjecture and the geometrization conjecture},}\ } (\bibinfo
  {year} {2006}),\ \Eprint {http://arxiv.org/abs/math/0612069v1 [math.DG]}
  {arXiv:math/0612069v1 [math.DG]} \BibitemShut {NoStop}%
\bibitem [{\citenamefont {Topping}(2006)}]{Topping06}%
  \BibitemOpen
  \bibfield  {author} {\bibinfo {author} {\bibfnamefont {P.}~\bibnamefont
  {Topping}},\ }\href@noop {} {\emph {\bibinfo {title} {Lecture Notes on
  {R}icci Flow}}}\ (\bibinfo  {publisher} {Cambridge University Press},\
  \bibinfo {address} {Cambridge},\ \bibinfo {year} {2006})\BibitemShut
  {NoStop}%
\bibitem [{\citenamefont {Lafon}(2004)}]{Lafon04}%
  \BibitemOpen
  \bibfield  {author} {\bibinfo {author} {\bibfnamefont {S.}~\bibnamefont
  {Lafon}},\ }\emph {\bibinfo {title} {Diffusion Maps and Geometric
  Harmonics}},\ \href@noop {} {Ph.D. thesis},\ \bibinfo  {school} {Yale
  University} (\bibinfo {year} {2004})\BibitemShut {NoStop}%
\bibitem [{\citenamefont {LeCun}\ \emph {et~al.}(1990)\citenamefont {LeCun},
  \citenamefont {Denker}, \citenamefont {Solla}, \citenamefont {Jackel},\ and\
  \citenamefont {Howard}}]{LeCunEtAl90}%
  \BibitemOpen
  \bibfield  {author} {\bibinfo {author} {\bibfnamefont {Y.}~\bibnamefont
  {LeCun}}, \bibinfo {author} {\bibfnamefont {J.~S.}\ \bibnamefont {Denker}},
  \bibinfo {author} {\bibfnamefont {S.~A.}\ \bibnamefont {Solla}}, \bibinfo
  {author} {\bibfnamefont {L.~D.}\ \bibnamefont {Jackel}}, \ and\ \bibinfo
  {author} {\bibfnamefont {R.~E.}\ \bibnamefont {Howard}},\ }in\ \href@noop {}
  {\emph {\bibinfo {booktitle} {Advances in Neural Information Processing
  Systems 2}}}\ (\bibinfo  {publisher} {Morgan Kaufmann Publishers Inc.},\
  \bibinfo {address} {Denver},\ \bibinfo {year} {1990})\ p.\ \bibinfo {pages}
  {598}\BibitemShut {NoStop}%
\bibitem [{\citenamefont {Cromer}\ and\ \citenamefont
  {Mann}(1968)}]{CromerMann68}%
  \BibitemOpen
  \bibfield  {author} {\bibinfo {author} {\bibfnamefont {D.~T.}\ \bibnamefont
  {Cromer}}\ and\ \bibinfo {author} {\bibfnamefont {J.~B.}\ \bibnamefont
  {Mann}},\ }\href@noop {} {\bibfield  {journal} {\bibinfo  {journal} {Acta
  Cryst. A}\ }\textbf {\bibinfo {volume} {24}},\ \bibinfo {pages} {321}
  (\bibinfo {year} {1968})}\BibitemShut {NoStop}%
\bibitem [{\citenamefont {Lovisolo}\ and\ \citenamefont
  {da~Silva}(2001)}]{LovisoloDaSilva01}%
  \BibitemOpen
  \bibfield  {author} {\bibinfo {author} {\bibfnamefont {L.}~\bibnamefont
  {Lovisolo}}\ and\ \bibinfo {author} {\bibfnamefont {E.~A.~B.}\ \bibnamefont
  {da~Silva}},\ }\href@noop {} {\bibfield  {journal} {\bibinfo  {journal} {IEEE
  Proc., Vis. Image Signal Process.}\ }\textbf {\bibinfo {volume} {148}},\
  \bibinfo {pages} {187} (\bibinfo {year} {2001})}\BibitemShut {NoStop}%
\bibitem [{Note1()}]{Note1}%
  \BibitemOpen
  \bibinfo {note} {The algorithm in Table~\ref {algNoiseFree} was executed with
  the following parameters: number of nearest neighbors in the sparse distance
  matrix $ d = 220 $; Gaussian kernel bandwidth $ \epsilon = 1 \times 10^4 $;
  number of datapoints for least-squares fitting $ r = 8 \times 10^4
  $.}\BibitemShut {Stop}%
\bibitem [{\citenamefont {Schwander}\ \emph
  {et~al.}(2011{\natexlab{b}})\citenamefont {Schwander} \emph
  {et~al.}}]{SchwanderEtAl10b}%
  \BibitemOpen
  \bibfield  {author} {\bibinfo {author} {\bibfnamefont {P.}~\bibnamefont
  {Schwander}} \emph {et~al.},\ }\href@noop {} {\enquote {\bibinfo {title}
  {Efficient interpolation of scattering data to an arbitrary grid},}\ }
  (\bibinfo {year} {2011}{\natexlab{b}}),\ \bibinfo {note} {to be
  published}\BibitemShut {NoStop}%
\bibitem [{\citenamefont {Marchesini}(2008)}]{Marchesini08}%
  \BibitemOpen
  \bibfield  {author} {\bibinfo {author} {\bibfnamefont {S.}~\bibnamefont
  {Marchesini}},\ }\href@noop {} {\enquote {\bibinfo {title} {Ab initio
  compressive phase retrieval},}\ } (\bibinfo {year} {2008}),\ \Eprint
  {http://arxiv.org/abs/0809.2006v1 [physics.optics]} {arXiv:0809.2006v1
  [physics.optics]} \BibitemShut {NoStop}%
\bibitem [{\citenamefont {Svensen}(1998)}]{Svensen98}%
  \BibitemOpen
  \bibfield  {author} {\bibinfo {author} {\bibfnamefont {J.~F.~M.}\
  \bibnamefont {Svensen}},\ }\emph {\bibinfo {title} {{GTM}: The Generative
  Topographic Mapping}},\ \href@noop {} {Ph.D. thesis},\ \bibinfo  {school}
  {Aston University} (\bibinfo {year} {1998})\BibitemShut {NoStop}%
\bibitem [{\citenamefont {Shneerson}\ \emph {et~al.}(2008)\citenamefont
  {Shneerson}, \citenamefont {Ourmazd},\ and\ \citenamefont
  {Saldin}}]{ShneersonEtAl08}%
  \BibitemOpen
  \bibfield  {author} {\bibinfo {author} {\bibfnamefont {V.~L.}\ \bibnamefont
  {Shneerson}}, \bibinfo {author} {\bibfnamefont {A.}~\bibnamefont {Ourmazd}},
  \ and\ \bibinfo {author} {\bibfnamefont {D.~K.}\ \bibnamefont {Saldin}},\
  }\href@noop {} {\bibfield  {journal} {\bibinfo  {journal} {Acta Cryst. A}\
  }\textbf {\bibinfo {volume} {64}},\ \bibinfo {pages} {303} (\bibinfo {year}
  {2008})}\BibitemShut {NoStop}%
\bibitem [{\citenamefont {Heisenberg}(1981)}]{Heisenberg81}%
  \BibitemOpen
  \bibfield  {author} {\bibinfo {author} {\bibfnamefont {W.}~\bibnamefont
  {Heisenberg}},\ }\href@noop {} {\emph {\bibinfo {title} {Der Teil und das
  Ganze}}}\ (\bibinfo  {publisher} {R. Piper \& Co. Verlag},\ \bibinfo
  {address} {Munich},\ \bibinfo {year} {1981})\ pp.\ \bibinfo {pages}
  {331--ff}\BibitemShut {NoStop}%
\bibitem [{\citenamefont {Saldin}\ \emph {et~al.}(2009)\citenamefont {Saldin},
  \citenamefont {Shneerson}, \citenamefont {Fung},\ and\ \citenamefont
  {Ourmazd}}]{SaldinEtAl09}%
  \BibitemOpen
  \bibfield  {author} {\bibinfo {author} {\bibfnamefont {D.}~\bibnamefont
  {Saldin}}, \bibinfo {author} {\bibfnamefont {V.~L.}\ \bibnamefont
  {Shneerson}}, \bibinfo {author} {\bibfnamefont {R.}~\bibnamefont {Fung}}, \
  and\ \bibinfo {author} {\bibfnamefont {A.}~\bibnamefont {Ourmazd}},\
  }\href@noop {} {\bibfield  {journal} {\bibinfo  {journal} {J. Phys.: Condens.
  Matter}\ }\textbf {\bibinfo {volume} {21}},\ \bibinfo {pages} {134014}
  (\bibinfo {year} {2009})}\BibitemShut {NoStop}%
\bibitem [{\citenamefont {Kostelec}\ and\ \citenamefont
  {Rockmore}(2003)}]{KostelecRockmore03}%
  \BibitemOpen
  \bibfield  {author} {\bibinfo {author} {\bibfnamefont {P.}~\bibnamefont
  {Kostelec}}\ and\ \bibinfo {author} {\bibfnamefont {D.}~\bibnamefont
  {Rockmore}},\ }in\ \href@noop {} {\emph {\bibinfo {booktitle} {Santa Fe
  Institute's Working Papers series}}}\ (\bibinfo {year} {2003})\ \bibinfo
  {note} {paper \#: 03-11-060}\BibitemShut {NoStop}%
\bibitem [{Note2()}]{Note2}%
  \BibitemOpen
  \bibinfo {note} {An explicit formula for the volume element associated with $
  B $ (the Haar measure), expressed in terms of the $ zyz $ Euler angles $ (
  \alpha ^1, \alpha ^2, \alpha ^3 ) $ parameterizing a rotation matrix $
  \protect \mathsf { R } $, is $ dV( \protect \mathsf { R } ) = \delimiter
  69640972 B \delimiter 86418188 ^{1/2} d\alpha ^1 \protect \tmspace
  +\thinmuskip {.1667em} d\alpha ^2 \protect \tmspace +\thinmuskip {.1667em}
  d\alpha ^3 $ with $ \delimiter 69640972 B \delimiter 86418188 = ( \protect
  \qopname \relax m{det}[ B_(\mu \nu )] )^{1/2} = \protect \qopname \relax
  o{sin}(\alpha ^2) $.}\BibitemShut {Stop}%
\end{thebibliography}

%

\end{document}